\documentclass[letter]{vldb} 

\usepackage{balance}  
\usepackage{algorithm}
\usepackage{epsfig}
\usepackage{graphicx}
\usepackage{esvect} 
\usepackage{xspace,amsmath,multirow,color,array,colortbl}
\usepackage{cite}
\usepackage{subfigure}
\usepackage[english]{babel}

\newcommand{\myhrule}{\rule[.5pt]{\hsize}{.5pt}}

\newcommand{\eat}[1]{}
\newcommand{\stab}{\rule{0pt}{8pt}\\[-1.6ex]}

\newcommand{\sstab}{\rule{0pt}{8pt}\\[-2.4ex]}

\newcommand{\bi}{\begin{itemize}}
\newcommand{\ei}{\end{itemize}}

\newcommand{\mat}[2]{{\begin{tabbing}\hspace{#1}\=\+\kill #2\end{tabbing}}}

\newcommand{\be}{\begin{enumerate}}
\newcommand{\ee}{\end{enumerate}}
\newcommand{\beqn}{\begin{eqnarray*}}
\newcommand{\eeqn}{\end{eqnarray*}}

\newcommand{\stitle}[1]{\vspace{1ex}\noindent{\bf #1}}

\newcommand{\ie}{\emph{i.e.,}\xspace}
\newcommand{\eg}{\emph{e.g.,}\xspace}
\newcommand{\wrt}{\emph{w.r.t.}\xspace}

\newcommand{\If}{\mbox{\bf if}\ }
\newcommand{\Let}{\mbox{\bf let}\ }

\newcommand{\Then}{\mbox{\bf then}\ }

\newcommand{\Else}{\mbox{\bf else}\ }

\newcommand{\While}{\mbox{\bf while}\ }

\newcommand{\Do}{\mbox{\bf do}\ }

\newcommand{\For}{\mbox{\bf for}\ }

\newcommand{\Or}{\mbox{\bf or}\ }
\renewcommand{\And}{\mbox{\bf and}\ }

\newcommand{\Break}{\mbox{\bf break}\ }

\newcommand{\Return}{\mbox{\bf return}\ }

\newcommand{\False}{\mbox{\em false}}
\newcommand{\True}{\mbox{\em true}}

\newcommand{\kw}[1]{{\ensuremath {\mathsf{#1}}}\xspace}

\newcounter{ccc}
\newcommand{\bcc}{\setcounter{ccc}{1}\theccc.}
\newcommand{\icc}{\addtocounter{ccc}{1}\theccc.}

\newcommand{\eop}{\hspace*{\fill}\mbox{$\Box$}}     
\newcounter{example}
\newenvironment{example}{
         \vspace{1.5ex}
         \refstepcounter{example}
         {\noindent\bf Example \theexample:}}{
         \eop\vspace{1.5ex}}

\renewcommand{\ni}{\noindent}

\newcommand{\nthesection}{\arabic{section}}
\newcounter{theorem}
\renewcommand{\thetheorem}{\arabic{theorem}}

\newenvironment{ttheorem}{\begin{em}
         \refstepcounter{theorem}
         {\vspace{1.5ex} \noindent\bf  Theorem  \thetheorem:}}{
        \end{em}\eop\vspace{1.5ex}} 
\newenvironment{pprop}{\begin{em}
        \refstepcounter{theorem}
        {\vspace{1.5ex}\noindent \bf Proposition \thetheorem:}}{
        \end{em}\eop\vspace{1.5ex}}
 \newenvironment{llemma}{\begin{em}
         \refstepcounter{theorem}
        {\vspace{1.5ex}\noindent\bf Lemma \thetheorem:}}{
         \end{em}\eop\vspace{1.5ex}} 
\newcounter{alg}[section]
\renewcommand{\thealg}{\nthesection.\arabic{alg}}

\newcounter{arule}
\renewcommand{\thearule}{\arabic{arule}}

\newcounter{claim}
\renewcommand{\theclaim}{\arabic{claim}}

\renewenvironment{proof}{
        \vspace{0ex}
        {\noindent\bf Proof:}}{\eop\vspace{1ex}}

\newcommand{\lsa}{\kw{LS}}
\newcommand{\dpa}{\kw{DP}}
\newcommand{\opwa}{\kw{OPW}}
\newcommand{\bqsa}{\kw{BQS}}
\newcommand{\fbqsa}{\kw{FBQS}}
\newcommand{\operaa}{\kw{OPERB}}
\newcommand{\operab}{\kw{OPERB}-\kw{A}}
\newcommand{\operb}{\kw{OPERB}}
\newcommand{\operba}{\kw{OPERB}-\kw{A}}

\newcommand{\taxi}{\kw{Taxi}}
\newcommand{\truck}{\kw{Truck}}

\newcommand{\sercar}{\kw{SerCar}}
\newcommand{\geolife}{\kw{GeoLife}}

\newcommand{\trajec}[1]{$\dddot{\mathcal{#1}}$}

\newcommand{\anoline}{\kw{AL}}

\title{One-Pass Error Bounded Trajectory Simplification}

\numberofauthors{1} 

\author{
\alignauthor
    Xuelian Lin\hspace{1.5ex} Shuai Ma$^{*}$\hspace{1.5ex} Han Zhang\hspace{1.5ex} Tianyu Wo\hspace{1.5ex} Jinpeng Huai\\
    \affaddr{SKLSDE Lab, Beihang University, Beijing, China}\\
    \affaddr{Beijing Advanced Innovation Center for Big Data and Brain Computing, Beijing, China}\\
    \email{\{linxl,  mashuai, zhanghan, woty, huaijp\}@buaa.edu.cn}
}

\begin{document}

\maketitle
\begin{abstract}
Nowadays, various sensors are collecting, storing and transmitting tremendous trajectory data, and it is known that raw trajectory data seriously wastes the storage, network band and computing resource. Line simplification (\lsa) algorithms are an effective approach to attacking this issue by compressing data points in a trajectory to a set of continuous line segments, and are commonly used in practice. However, existing  \lsa algorithms are not sufficient for the needs of sensors in mobile devices. In this study, we first develop a one-pass error bounded trajectory simplification algorithm (\operb), which scans each data point in a trajectory once and only once. We then propose an aggressive one-pass error bounded trajectory simplification algorithm (\operba), which allows interpolating new data points into a trajectory under certain conditions. Finally, we experimentally verify that our approaches (\operb and \operba) are both efficient and effective, using four real-life trajectory datasets.
\end{abstract}

%

%

%
%



\section{Introduction}
\label{sec-into}
\begin{figure}[tb!]\label{fig:dp}
\centering
\includegraphics[scale=0.64]{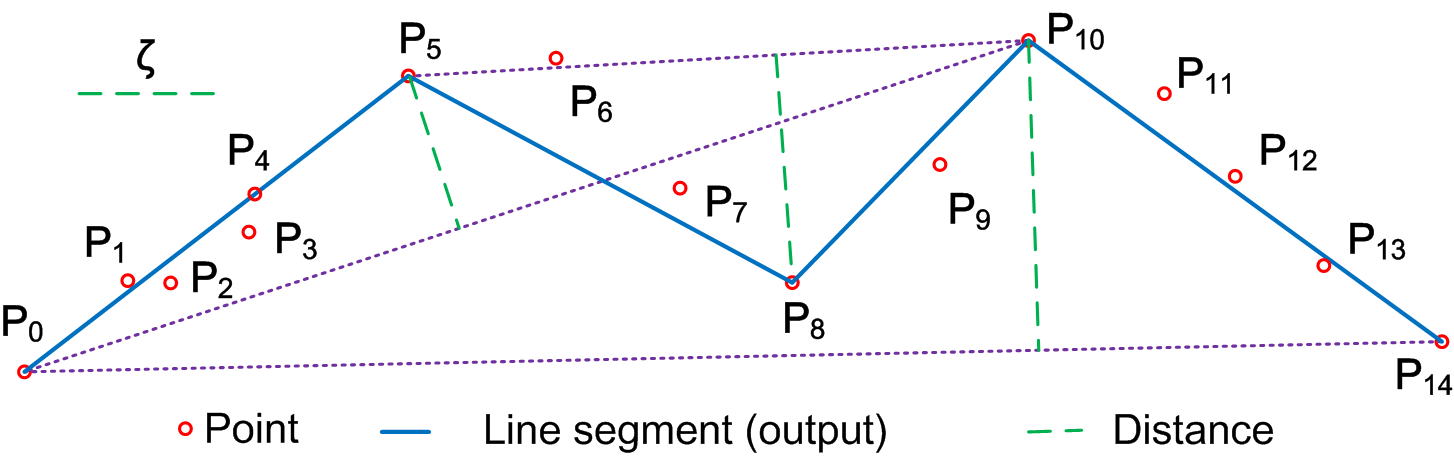}
\vspace{-1.5ex}
\caption{\small A trajectory $\dddot{\mathcal{T}}[P_0, \ldots, P_{14}]$  with fifteen points is represented by four continuous line segments (solid blue), compressed by the Douglas--Peucker algorithm \cite{Douglas:Peucker}.}
\vspace{-4ex}
\end{figure}

Various mobile devices, such as smart-phones, on-board diagnostics, \eat{personal navigation devices,} and wearable smart devices, have been widely using their sensors to collect massive trajectory data of moving objects at a certain sampling rate (\eg 5 seconds), and transmit it to cloud servers for location based services, trajectory mining and many other applications.
It is known that transmitting and storing raw trajectory data consumes too much network bandwidth and storage capacity \cite{Meratnia:Spatiotemporal,Keogh:online,Liu:BQS, Muckell:Compression, Chen:Trajectory, Kaul:predicting, Chen:Fast,Cao:Spatio, Popa:Spatio, Schmid:Semantic,Richter:Semantic,Long:Direction,Nibali:Trajic}.
Further, we find that the online transmitting of raw trajectories also seriously aggravates several other issues such as out-of-order and duplicate data points in our experiences when implementing an online vehicle-to-cloud data transmission system.
Fortunately, these issues can be resolved or greatly alleviated by the trajectory compression techniques \cite{Douglas:Peucker, Hershberger:Speeding, Meratnia:Spatiotemporal,Liu:BQS,Muckell:Compression,Chen:Trajectory, Chen:Fast,Keogh:online, Cao:Spatio, Shi:Survey, Richter:Semantic,Long:Direction,Song:PRESS
,Nibali:Trajic}, among which line simplification based methods are widely used \cite{Douglas:Peucker, Hershberger:Speeding, Keogh:online,Liu:BQS, Muckell:Compression, Chen:Trajectory, Chen:Fast, Cao:Spatio, Shi:Survey}, due to their distinct advantages: (a) simple and easy to implement, (b) no need of extra knowledge and suitable for freely  moving  objects \cite{Popa:Spatio}, and (c) bounded errors with good compression ratios. Line simplification algorithms belong to lossy compression, and use a set of continuous line segments to represent a compressed trajectory, as shown in Figure~\ref{fig:dp}.

\eat{The sensors usually have a limited memory, network band and computing ability, and  hence, a fast trajectory compression algorithm with a high compression ratio means less computing resource consumption.}

The most notable line simplification (\lsa) algorithm is the Douglas-Peucker algorithm \cite{Douglas:Peucker} invented in 1970s, for reducing the number of points required to represent a digitized line or its caricature in the context of computer graphics and image processing. The basic Douglas-Peucker algorithm (\dpa) is a batch method with a time complexity of $O(n^2)$, where $n$ is the number of data points in a given trajectory to be compressed. Its batch nature and high time complexity make it not suitable for the online scenarios. Several \lsa algorithms have been developed based on \dpa, \eg by combining \dpa with sliding/open windows \cite{Meratnia:Spatiotemporal, Keogh:online} for online processing. However, these methods still have a high time and/or space complexity, which significantly hinders their utility in resource-constrained mobile devices \cite{Liu:BQS}.

Recently, \bqsa \cite{Liu:BQS} has been proposed, using a new distance checking method by picking out at most eight special points from an open window based on a convex hull, \eg a rectangular bounding box with two bounding lines, so that when a new point is added to a window, it only needs to calculate the distances of the  special points to a line, instead of all data points in the window, in many cases. The time complexity of \bqsa remains $O(n^2)$ in the worst case, as \bqsa falls back to \dpa when the eight special points cannot be used. However, its simplified version, \fbqsa directly outputs a line segment, and starts a new window when the eight special points cannot bound all the points considered so far. Indeed, \fbqsa has a linear time complexity, and is the fastest \lsa based solution for trajectory compression.


An \lsa algorithm is {\em one-pass} if it processes each point in a trajectory once and only once when compressing the trajectory.
Obviously, one-pass algorithms have low time and space complexities, and are more appropriate for online processing.
Unfortunately, existing algorithms such as \dpa, \bqsa and \fbqsa are not one-pass, as data points in a trajectory are processed multiple times in these algorithms. Indeed, {\em it remains open whether there exist one-pass error bounded \lsa based effective trajectory compression algorithms}.



\stitle{Contributions \& Roadmap}.
To this end, we propose two one-pass error bounded \lsa based algorithms for compressing trajectories in an efficient and effective way.

\stab (1) We first develop a one-pass error bounded trajectory simplification algorithm (\operb, Section \ref{sec-operb}) that runs in $O(n)$ time and $O(1)$ space. \operb is based on a novel local distance checking method, and equipped with five optimization techniques to further improve its compression ratio.

\stab (2) We then propose an aggressive one-pass error bounded trajectory simplification algorithm (\operab, Section \ref{sec-operba-agg}) that remains in $O(n)$ time and $O(1)$ space.
\operba allows interpolating new data points into a trajectory under certain conditions and with practical considerations. The rational lies in that moving objects  have sudden track changes while data points may  not be sampled due to various reasons.

\stab (3) Using four real-life trajectory datasets (\taxi, \truck, \sercar, \geolife), we finally conduct an extensive experimental study (Section \ref{sec-exp}), by comparing our algorithms \operaa and \operab with \fbqsa (\emph{the fastest} existing \lsa algorithm) and \dpa (the {\em best} existing \lsa algorithm in terms of {\em compression ratio}). We find that \operaa and \operab are on average $(4.1, 4.1, 5.4, 5.2)$ times faster than \fbqsa on (\taxi, \truck, \sercar, \geolife), respectively.
For compression ratios, \operaa is comparable with \dpa, and \operab is better than \dpa that is on average ($84.2\%, 86.4\%, 97.1\%, 94.7\%$) of \dpa on (\taxi, \truck, \sercar, \geolife), respectively.



An extended version and some used datasets  are at~\cite{OPERB-full}.

\section{Related Work}
Trajectory compression algorithms are normally classified into two categories, namely lossless compression and lossy compression\cite{Muckell:Compression}.
(1) Lossless compression methods enable exact reconstruction of the original data from the compressed data without information loss. For example, delta compression \cite{Nibali:Trajic} is a lossless compression technique, which has zero error and a time complexity of $O(n)$, where $n$ is the number of data points in a trajectory. The limitation of lossless compression lies in that its compression ratio is relatively poor \cite{Nibali:Trajic}.
(2) In contrast, lossy compression methods allow errors or derivations, compared with the original trajectories.
These techniques typically identify important data points, and remove statistical redundant data points from a trajectory, or replace original data points in a trajectory with other places of interests, such as roads and shops.
They focus on good compression ratios with acceptable errors, and a large number of lossy trajectory compression techniques have been developed.
In this work we focus on lossy compression of trajectory data,

We next introduce the related work on lossy trajectory compression  from two aspects: line simplification based methods and semantics based methods.

\stitle{Line simplification based methods}.
Line simplification based methods not only have good compression ratios and deterministic error bounds, but also are easy to implement (see an evaluation report \cite{Shi:Survey})). Hence, they are widely used in practice, even for freely moving objects without the restriction of road networks.  And according to the way they process trajectories, they are further divided into batch processing and online processing methods~\cite{Nibali:Trajic}.

\eat{
Line simplification based methods not only have good compression ratios and deterministic error bounds, but also are easy to implement. Hence, they are widely used in practice, even for freely moving objects without the restriction of road networks.
The error bounds are typically calculated by the Euclidean distance, but there are variants of the Euclidean distance, such as the synchronous Euclidean distance \cite{Meratnia:Spatiotemporal}. 
}


\sstab (1) Batch algorithms require that all trajectory points must be loaded before they start compressing. Batch algorithms can be either top-down or bottom-up.
Top-down algorithms recursively divide a trajectory into sub-trajectories until the stopping condition is met\cite{Keogh:online}.
The \dpa algorithm \cite{Douglas:Peucker} is the most classic top-down approach, and \cite{Meratnia:Spatiotemporal} improves \dpa with the synchronous Euclidean distance, instead of the Euclidean distance.
Bottom-up algorithms~\cite{Keogh:online, Chen:Fast} are the natural complement to top-down ones, which recursively merge adjacent sub-trajectories with the smallest distance, initially $n/2$  sub-trajectories for a trajectory with $n$ points, until the stopping condition is met. Note that the distances of newly generated line segments are recalculated during the process.

\sstab(2) Online algorithms do not need to have the entire trajectory ready before they start compressing, and are appropriate for compressing trajectories on sensors of mobile devices. Existing online algorithms~\cite{Keogh:online,Meratnia:Spatiotemporal,Muckell:Compression} usually use a fixed or open window and compress sub-trajectories in the window.

However, these existing online algorithms are not one-pass.  In this study, we propose a novel local distance checking method, based on which we develop one-pass online algorithms that are totally different from the window based algorithms. Further, as shown in the experimental study, our approaches are clearly superior to the existing online algorithms, in terms of both efficiency and effectiveness.

\stitle{Semantics based methods}.
The trajectories of certain moving objects such as cars and trucks are constrained by road networks. These moving objects typically travel along road networks, instead of the line segment between two points. Trajectory compression methods based on road networks \cite{Chen:Trajectory, Popa:Spatio,Civilis:Techniques,Hung:Clustering, Gotsman:Compaction, Song:PRESS}  project trajectory points onto roads (also known as Map-Matching). Moreover, \cite{Gotsman:Compaction, Song:PRESS} mines and uses high frequency patterns of compressed trajectories, instead of roads, to further improve compression effectiveness.
Some methods \cite{Schmid:Semantic, Richter:Semantic} compress trajectories beyond the use of road networks, which further make use
of other user specified domain knowledge, such as places of interests along the trajectories \cite{Richter:Semantic}.
There are also compression algorithms preserving the direction of the trajectory \cite{Long:Direction, Long:Trajectory}. 

These approaches are orthogonal to line simplification based methods, and may be combined with each other to further improve the effectiveness of trajectory compression.



\eat{
\stitle{Lossless compression methods}. The above line implication methods and semantics based methods belong to lossy compression techniques.
Delta compression \cite{Nibali:Trajic} is a lossless compression, which has a zero error bound and a time complexity of $O(n)$, where $n$ is the number of data points in a trajectory. However, its compression ratio is relatively poor \cite{Nibali:Trajic}. In contrast, our algorithms \operb and \operba proposed in this study not only run in  $O(n)$ time, but also have a very good compression ratio.
}

\vspace{-0.5ex}
\section{Preliminaries}
\label{sec-background}

In this section, we introduce some basic concepts and existing algorithms for trajectory simplification.
\subsection{Basic Notations}

\newdef{definition}{Definition}
\newtheorem{theorem}{Theorem}
\newtheorem{lemma}{Lemma}


\stitle{Points ($P$)}. A data point is defined as a triple $P(x, y, t)$, which represents that a moving object is located at {\em longitude} $x$ and {\em latitude} $y$ at {\em time} $t$. Note that data points can be viewed as points in a three-dimension Euclidean space.

\stitle{Trajectories ($\dddot{\mathcal{T}}$)}. A trajectory $\dddot{\mathcal{T}}[P_0, \ldots, P_n]$ is a sequence of data points in a monotonically increasing order of their associated time values (\ie $P_i.t < P_j.t$ for any $0\le i<j\le n$). Intuitively, a trajectory is the path (or track) that a moving object follows through space as a function of time~\cite{physics-trajectory}.

\stitle{Directed line segments ($\mathcal{L}$)}. A directed line segment (or line segment for simplicity) $\mathcal{L}$ is defined as $\vv{P_{s}P_{e}}$, which represents the closed line segment that connects the start point $P_s$ and the end point $P_e$.
Note that here $P_s$ or $P_e$ may not be a point in a trajectory $\dddot{\mathcal{T}}$, and hence, we also use notation $\mathcal{R}$ instead of $\mathcal{L}$ when both $P_s$ and $P_e$ belong to $\dddot{\mathcal{T}}$.

\eat{
\textcolor[rgb]{0.00,0.00,1.00}{Note that in this definition, $P_s$ or $P_e$ could be a point not belonging to any $\dddot{\mathcal{T}}$. In the rest of the paper, we may use notation $\mathcal{R}$ instead of $\mathcal{L}$ for a directed line segment whose $P_s$ and $P_e$ are both belonging to the same trajectory $\dddot{\mathcal{T}}$.}
}

We also use $|\mathcal{L}|$ and $\mathcal{L}.\theta\in [0, 2\pi)$ to denote the length of a directed line segment $\mathcal{L}$, and its angle with the $x$-axis of the coordinate system $(x, y)$, where $x$ and $y$ are the longitude and latitude, respectively.
That is, a directed line segment $\mathcal{L}$ = $\vv{P_{s}P_{e}}$ can be treated as a triple $(P_s, |\mathcal{L}|, \mathcal{L}.\theta)$.

\stitle{Piecewise line representation ($\overline{\mathcal{T}}$)}. A piece-wise line representation of a trajectory $\dddot{\mathcal{T}}[P_0, \ldots, P_n]$ is $\overline{\mathcal{T}}[\mathcal{L}_0, \ldots , \mathcal{L}_m]$ ($0< m \le n$), a sequence of continuous directed line segments $\mathcal{L}_{i}$ = $\vv{P_{s_i}P_{e_i}}$ of $\dddot{\mathcal{T}}$ ($i\in[0,m]$)  such that $\mathcal{L}_{0}.P_{s_0} = P_0$, $\mathcal{L}_{m}.P_{e_m} = P_n$ and  $\mathcal{L}_{i}.P_{e_i}$ = $\mathcal{L}_{i+1}.P_{s_{i+1}}$ for all $i\in[0, m-1]$. Note that each directed line segment in $\overline{\mathcal{T}}$ essentially represents a continuous sequence of data points in $\dddot{\mathcal{T}}$.

\stitle{Included angles ($\angle$)}. Given two directed line segments $\mathcal{L}_1$ = $\vv{P_{s}P_{e_1}}$ and $\mathcal{L}_2$ = $\vv{P_{s}P_{e_2}}$ with the same start point $P_s$, the included angle from $\mathcal{L}_1$ to $\mathcal{L}_2$, denoted as $\angle(\mathcal{L}_1, \mathcal{L}_2)$,  is $\mathcal{L}_2.\theta - \mathcal{L}_1.\theta$. For convenience, we also represent the included angle  $\angle(\mathcal{L}_1, \mathcal{L}_2)$ as $\angle{P_{e_1}P_sP_{e_2}}$.

\stitle{Distances ($d$)}. Given a point $P_i$ and a directed line segment $\mathcal{L}$ = $\vv{P_{s}P_{e}}$, the distance of $P_i$ to $\mathcal{L}$, denoted as $d(P_i, \mathcal{L})$, is the Euclidean distance from $P_i$ to the line $\overline{P_sP_e}$, commonly adopted by most existing \lsa methods, \eg~\cite{Douglas:Peucker, Hershberger:Speeding, Keogh:online, Chen:Fast, Liu:BQS}.

\begin{figure}[tb!]
\centering
\includegraphics[scale = 0.76]{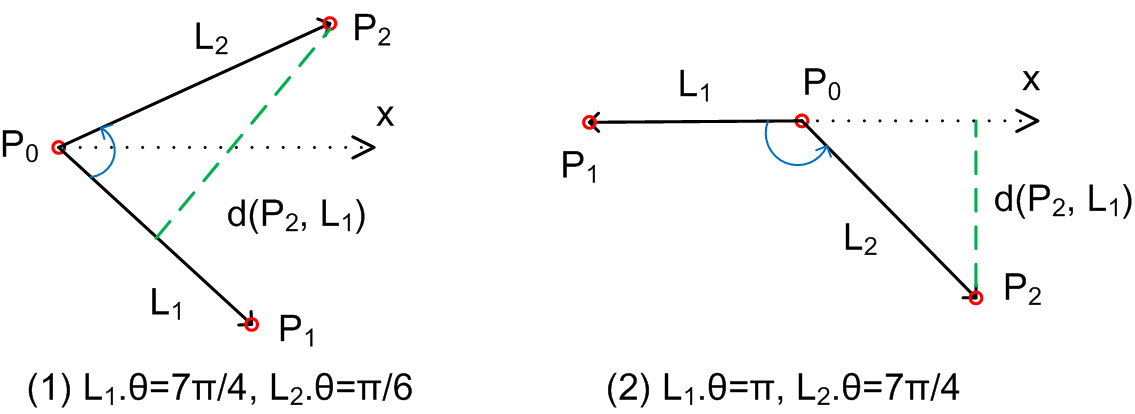}
\vspace{-2ex}
\caption{\small Example included angles and distances.}
\label{fig:distance}
\vspace{-3.5ex}
\end{figure}

\begin{example}
\label{exm-notations}
\eat{(1) Recall that Figure~\ref{fig:dp} shows that a \textcolor[rgb]{0.00,0.00,1.00}{trajectory $\dddot{\mathcal{T}}[P_0$, $\ldots, P_{14}]$} with fifteen points.
The four continuous directed line segments $\{\vv{P_0P_5}$, $\vv{P_5P_8}$, $\vv{P_8P_{10}}$, $\vv{P_{10}P_{14}}\}$ forms a piecewise line representation of $\dddot{\mathcal{T}}$.}
(1) In Figure~\ref{fig:dp}, the four continuous directed line segments $\vv{P_0P_5}$, $\vv{P_5P_8}$, $\vv{P_8P_{10}}$, $\vv{P_{10}P_{14}}$ form a piecewise line representation of trajectory $\dddot{\mathcal{T}}[P_0$, $\ldots, P_{14}]$.

\sstab(2) Figure~\ref{fig:distance} shows two different cases of included angles. In each case, there are two directed line segments $\mathcal{L}_1= \vv{P_0P_1}$ and $\mathcal{L}_2=\vv{P_0P_2}$ with the same start point $P_0$. The included angle $\angle(\mathcal{L}_1, \mathcal{L}_2)$ from $\mathcal{L}_1$  to $\mathcal{L}_2$  lies in $(-2\pi, 2\pi)$, and is $\frac{-19\pi}{12}$ and $\frac{3\pi}{4}$ in Figures~\ref{fig:distance}(1)\&(2), respectively.

\sstab(3) The distance $d(P_2, \mathcal{L}_1)$ is illustrated in Figures~\ref{fig:distance} as dotted green line segments.
\end{example}
\vspace{-3ex}

\subsection{Line Simplification Algorithms}
\label{subsec-lsalg}

Line simplification (\lsa) algorithms are a type of important and widely adopted trajectory compression methods, and we next briefly introduce these algorithms.

\eat{
\begin{algorithm}[tb!]
\begin{small}
\caption{: Basic Douglas-Peucker Algorithm}\label{alg:dp}
$\dpa(\dddot{\mathcal{T}}[P_0,\ldots,P_n], \zeta)$\\
1	~\For each point $P_i$ ($i\in[0,n]$) in $\dddot{\mathcal{T}}[P_0, \ldots, P_n]$ \Do\\
2	\hspace{4ex}compute $d(P_i, {\mathcal{L}})$ between $P_i$ and ${\mathcal{L}}(P_0, P_n)$;\\
3	~\Let $d(P_k, {\mathcal{L}})$ := $\max \{d(P_0, {\mathcal{L}}), \ldots, d(P_n, {\mathcal{L}}) \}$;\\
4	~\If $d(P_k, {\mathcal{L}}) <\zeta$ \Then \\
5	\hspace{4ex}\Return $\{\mathcal{L}(P_0,P_n)\}$.\\
6	~\Else\\
7	\hspace{4ex}\Return $\dpa($\trajec{T}$[P_0, \ldots, P_k], \zeta)\cup\dpa($\trajec{T}$[P_{k+1}, \ldots, P_n], \zeta)$.
\end{small}
\end{algorithm}
}

\stitle{Basic Douglas-Peucker algorithm}. We first introduce the Basic Douglas-Peucker (\dpa) algorithm  \cite{Douglas:Peucker} shown in Figure~\ref{alg:dp}, the foundation of many subsequent \lsa algorithms.

Given a trajectory $\dddot{\mathcal{T}}[P_0, \ldots, P_n]$ and an error bound $\zeta$, algorithm \dpa uses the first point $P_0$ and the last point $P_n$ of \trajec{T} as the start point $P_s$ and the end point $P_e$ of the first line segment $\mathcal{L}(P_0, P_n)$, then it calculates the distance $d(P_i, {\mathcal{L}})$ for each point $P_i$ ($i\in[0,n]$) (lines 1--2). If $d(P_k, {\mathcal{L}})$ = $\max \{d(P_0, {\mathcal{L}}), \ldots, d(P_n, {\mathcal{L}}) \} \le \zeta$, then it returns $\{\mathcal{L}(P_0,P_n)\}$ (lines 3--5). Otherwise, it divides \trajec{T} into two sub-trajectories \trajec{T}$[P_0, \ldots, P_k]$ and \trajec{T}$[P_{k}, \ldots, P_n]$, and recursively compresses these sub-trajectories until the entire trajectory has been considered (lines 6--7).

The \dpa algorithm is clearly a batch algorithm, as the entire trajectory is needed at the beginning \cite{Meratnia:Spatiotemporal}, and its time complexity is $O(n^2)$. Moreover, \cite{Hershberger:Speeding} developed an improved method with a time complexity of $O(n\log n)$.

\begin{figure}[tb!]
\begin{center}
{\small
\begin{minipage}{3.36in}
\myhrule \vspace{-2ex}
\mat{0ex}{
{\bf Algorithm}~$\dpa(\dddot{\mathcal{T}}[P_0,\ldots,P_n], \zeta)$\\
\sstab
\bcc \hspace{1ex}\=\For each point $P_i$ ($i\in[0,n]$) in $\dddot{\mathcal{T}}[P_0, \ldots, P_n]$ \Do\\
\icc \>\hspace{3ex}compute $d(P_i, {\mathcal{L}})$ between $P_i$ and ${\mathcal{L}}(P_0, P_n)$;\\
\icc \> \Let $d(P_k, {\mathcal{L}})$ := $\max \{d(P_0, {\mathcal{L}}), \ldots, d(P_n, {\mathcal{L}}) \}$;\\
\icc \> \If $d(P_k, {\mathcal{L}}) \le\zeta$ \Then \\
\icc \> \hspace{3ex}\Return $\{\mathcal{L}(P_0,P_n)\}$.\\
\icc \> \Else\\
\icc \> \hspace{3ex}\Return $\dpa($\trajec{T}$[P_0, \ldots, P_k], \zeta)\cup\dpa($\trajec{T}$[P_{k}, \ldots, P_n], \zeta)$.
}
\vspace{-2.5ex}
\myhrule
\end{minipage}
}
\end{center}
\vspace{-3ex}
\caption{\small Basic Douglas-Peucker algorithm}\label{alg:dp}
\vspace{-4ex}
\end{figure}

\begin{example}
\label{exm-alg-lsa}
Consider the trajectory $\dddot{\mathcal{T}}[P_0,\ldots,P_{14}]$ shown in Figure~\ref{fig:dp}. Algorithm $\dpa$ firstly creates $\vv{P_0P_{14}}$, then it calculates the distance of each point in $\{P_0,\ldots,P_{14}\}$ to $\vv{P_0P_{14}}$. It finds that $P_{10}$ has the maximum distance to $\vv{P_0P_{14}}$, which is greater than $\zeta$. Then it goes to compress sub-trajectories $[P_0, \ldots, P_{10}]$ and $[P_{10}, \ldots, P_{14}]$, separately. Similarly, sub-trajectory $[P_0,\ldots, P_{10}]$ is split to $[P_0$, $\ldots$, $P_5]$ and $[P_5$,  $\ldots$, $P_{10}]$, and $[P_5$, $\ldots$, $P_{10}]$ is split to $[P_5, \ldots, P_8]$ and $[P_8$, $P_9$, $P_{10}]$. Finally, algorithm $\dpa$ outputs four continuous directed line segments $\{\vv{P_0P_5}$, $\vv{P_5P_8}$, $\vv{P_8P_{10}}$, $\vv{P_{10}P_{14}}\}$, \ie a piece-wise line representation of trajectory $\dddot{\mathcal{T}}[P_0,\ldots,P_{14}]$
\end{example}

\vspace{-1ex}
\stitle{Online algorithms}. We next introduce two classes of \dpa based online algorithms that make use of sliding windows to speed up the compressing efficiency \cite{Meratnia:Spatiotemporal,Keogh:online,Liu:BQS}.

Given a trajectory $\dddot{\mathcal{T}}[P_0, \ldots, P_n]$ and an error bound $\zeta$, algorithm \opwa~\cite{Meratnia:Spatiotemporal} maintains a window $W[P_s, \ldots, P_k]$, where $P_s$ and $P_k$ are the start and end points, respectively. Initially, $P_s$ = $P_0$ and $P_k$ = $P_1$, and the window $W$ is gradually expanded by adding new points one by one. \opwa tries to compress all points in $W[P_s, \ldots, P_k]$ to a single line segment $\mathcal{L}(P_{s}, P_{k})$. If the distances $d(P_i, {\mathcal{L}})\le \zeta$ for all points $P_i$ ($i\in[s, k]$), it simply expands $W$ to $[P_s, \ldots, P_k, P_{k+1}]$ $(k+1\le n)$ by adding a new point $P_{k+1}$. Otherwise, it produces a new line segment $\mathcal{L}(P_{s}, P_{k-1})$, and replaces $W$ with a new window $[P_{k-1},\ldots,P_{k+1}]$. The above process repeats until all points in $\dddot{\mathcal{T}}$ have been considered.  Algorithm \opwa is not efficient enough for compressing long trajectories as it remains in $O(n^2)$ time, the same as the \dpa algorithm.

\bqsa \cite{Liu:BQS} reduces the compression time by introducing a convex hull that bounds a certain number of points. For a buffered sub trajectory $[P_s, \ldots, P_k]$, it splits the space into four quadrants. For each quadrant, a rectangular bounding box is firstly created using the least and highest $x$ values and the least and highest $y$ values among points $P_s,\ldots,P_k$. Then another two bounding lines connecting points $P_s$ and $P_{h}$ and points $P_s$ and $P_{l}$ are created such that lines $\overline{P_sP_{h}}$ and $\overline{P_sP_{l}}$ have the largest and smallest angles with the $x$-axis, respectively. Here $P_{h},P_{l} \in\{P_s,\ldots,P_k\}$. The bounding box and the two lines together form a convex hull.
\bqsa picks out at most eight significant points in a quadrant. 
In many cases, (1) it only calculates the distances of the significant points to line $\overline{P_sP_k}$;
otherwise, (2) it needs to compute all distances $d(P_i, {\mathcal{L}(P_s,P_k)})$ ($i\in[s, k]$) as \dpa.
\bqsa remains in $O(n^2)$ time.
However, its simplified version \fbqsa essentially avoids case (2) to achieve an $O(n)$ time complexity.

\begin{figure}[tb!]
\centering
\includegraphics[scale = 0.62]{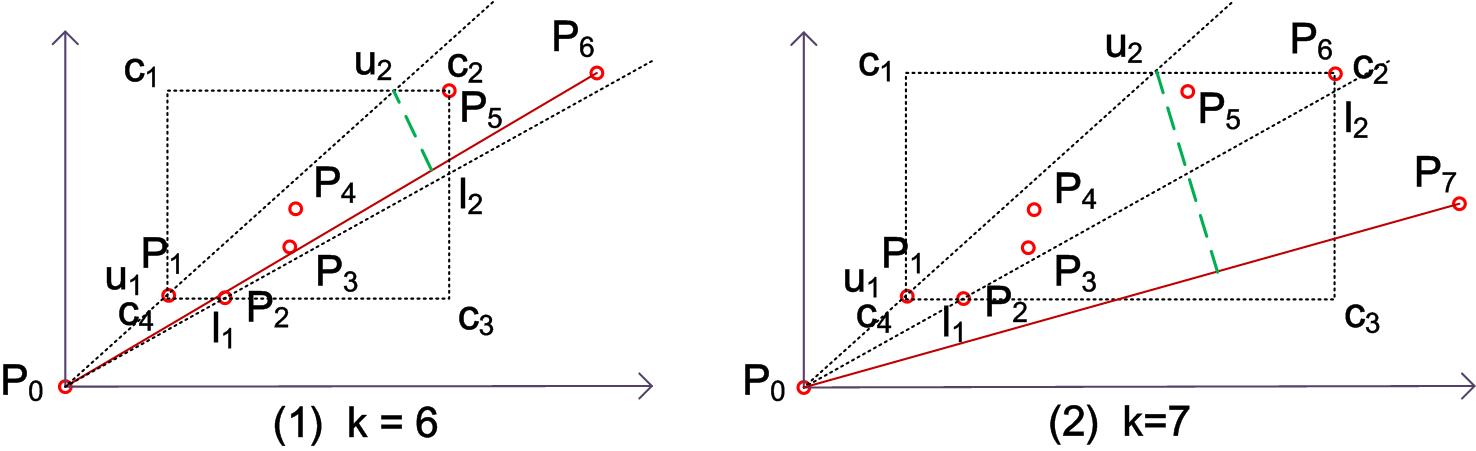}
\vspace{-2ex}
\caption{{\small Examples for algorithm \bqsa.}}
\label{fig:bqs}
\vspace{-3ex}
\end{figure}

\begin{example}
\label{exm-alg-bqs}
In Figure~\ref{fig:bqs}, the bounding box $c_1c_2c_3c_4$ and the two lines $\overline{P_sP_{h}} = \overline{P_0P_1}$ and $\overline{P_sP_{l}} = \overline{P_0P_2}$ form a convex hull $u_1u_2c_2l_2l_1c_4$. \bqsa computes the distances of $u_1,u_2,c_2,l_2,l_1$ and $c_4$ to line $\overline{P_0P_6}$ when $k=6$ or to line $\overline{P_0P_7}$ when $k=7$.

When $k=6$, all these distances to $\overline{P_0P_6}$  are less than $\zeta$, hence \bqsa goes on to the next point (case 1); When $k=7$,
the max and min distances to $\overline{P_0P_7}$ are larger and less than $\zeta$, respectively, and \bqsa needs to compress sub-trajectory $[P_0, \ldots, P_7]$ along the same line as \dpa (case 2).
\end{example}

\vspace{-1ex}
\stitle{Error bounded algorithms}. Given a trajectory \trajec{T} and its compression  algorithm $\mathcal{A}$ that produces another trajectory \trajec{T'},
we say that algorithm $\mathcal{A}$ is error bounded by $\zeta$ if  for each point $P$ in \trajec{T}, there exists a point $P_j$ in \trajec{T'} with $d(P, \mathcal{L}(P_j,P_{j+1}))\le \zeta$.
Note that all the above \lsa algorithms are error bounded by $\zeta$, a parameter typically set by experts based on the need and analysis of applications.

\vspace{-0.5ex}
\section{One-Pass Simplification}
\label{sec-operb}

In this section, we first develop a local distance checking approach that is the key for one-pass trajectory simplification algorithms. We then present a \underline{O}ne-\underline{P}ass \underline{ER}ror \underline{B}ounded trajectory simplification algorithm, referred to as  \operb. Finally, we propose five optimization techniques.

\subsection{Local Distance Checking}
\label{subsec-fittingfunction}

Existing trajectory simplification algorithms (\eg \ \dpa  \cite{Douglas:Peucker} and online algorithms \cite{Meratnia:Spatiotemporal,Keogh:online, Liu:BQS}) essentially employ a global distance checking approach to assuring error bounds, although online algorithms restrict the checking within a window. That is to say, whenever a new directed line segment $\mathcal{R}_i$ = $\vv{P_sP_{s+i}}$ $(i\in [1, k])$ is formed for a sub-trajectory $\dddot{\mathcal{T}_s}[P_s, \ldots, P_{s+k}]$, these algorithms always check its distances to all or a subset of data points $\{P_{s}, \ldots, P_{s+i}\}$ to $\mathcal{R}_i$, and, therefore, a data point is checked multiple times, depending on its order in the trajectory and the number of directed line segments formed. Hence, {\em an appropriate local distance checking approach is needed in the first place for designing one-pass trajectory simplification algorithms}.

Consider an error bound $\zeta$ and a sub-trajectory $\dddot{\mathcal{T}_s}[P_s,$ $\ldots, P_{s+k}]$. To achieve the local distance checking, \operb first dynamically maintains a directed line segment $\mathcal{L}_i$ ($i\in[1,k]$), whose start point is fixed with $P_s$ and its end point is identified (may not in $\{P_s, \ldots, P_{s+i}\}$) to {\em fit} all the previously processed points $\{P_s, \ldots, P_{s+i}\}$. The directed line segment $\mathcal{L}_i$ is built by a function named \emph{fitting function $\mathbb{F}$}, such that when a new point $P_{s+i+1}$ is considered, only its distance to the directed line segment $\mathcal{L}_i$ is checked, instead of checking the distances of all or a subset of data points of $\{P_{s}, \ldots, P_{s+i}\}$ to $\mathcal{R}_{i+1}$ = $\vv{P_sP_{s+i+1}}$  as the global distance checking does. In this way, a data point is checked only once during the entire process of trajectory simplification.

\eat{
More specifically, the state-of-the-art of the local distance checking approach can be demonstrated as following.

\vspace{1.5ex}

{\em The property ``$d(P_{s+i}, {\vv{P_{s}P_{s+k}})}\le \zeta$ for all points $P_{s+i}$ ($i\in[0, k]$) of $\dddot{\mathcal{T}_s}$", where $\vv{P_{s}P_{s+k}}$ is the line segment connecting the start point $P_{s}$ and the end point $P_{s+k}$, will be hold if

    (1.1) $\mathcal{L}_0$  = $\vv{P_sP_{s}}$ (\ie $|\mathcal{L}_0|$ = 0 and $\mathcal{L}_0.\theta$ = 0),

    (1.2) $\mathcal{L}_{i}$ = $\mathbb{F}(P_{s+i}, \mathcal{L}_{i-1})$ for $i\in[1, k+1]$, and

    (2.1) $d(P_{s+i}, \mathcal{L}_{i-1})\le \zeta/2$ for each $i\in[0, k]$.
}

\vspace{1.5ex}

\ni Here (1.1) and (1.2) are used to maintain the directed line segment $\mathcal{L}_i$ ($i\in[1,k]$), where $\mathbb{F}$ is referred to as a \emph{fitting function} which dynamically adjusts the  directed line segment $\mathcal{L}_i$, and (2.1) is the distance checking.
}

We next present the details of our fitting function   $\mathbb{F}$ that is designed for local distance checking.

\stitle{Fitting function $\mathbb{F}$}. Given an error bound $\zeta$ and a sub-trajectory $\dddot{\mathcal{T}_s}[P_s, \ldots, P_{s+k}]$, $\mathbb{F}$ is as follows.

\begin{small}
\vspace{-2ex}
\begin{equation*}
\label{equ-function}
\hspace{-1.5ex}\left\{
    \begin{aligned}
        &\hspace{-1.5ex}\left[
            \begin{aligned}
            & \mathcal{L}_{i} = \mathcal{L}_{i-1}\\
            \end{aligned}
        \right]\hspace{12.5ex}~when~(|\mathcal{R}_{i}| - |\mathcal{L}_{i-1}|) \le \frac{\zeta}{4} \hspace{14.5ex}(1)\\
        &\hspace{-1.5ex}\left[
            \begin{aligned}
            & |\mathcal{L}_{i}|  = j*{\zeta}/{2} \\
            & \mathcal{L}_{i}.\theta = \mathcal{R}_{i}.\theta    \\
            \end{aligned}
        \right]\hspace{8.5ex}~when~|\mathcal{R}_{i}| >  \frac{\zeta}{4}~\And~|\mathcal{L}_{i-1}|=0   \hspace{8ex}(2)~ \\
        &\hspace{-1.5ex}\left[
            \begin{aligned}
            & |\mathcal{L}_{i}|  = j*{\zeta}/{2}\\
            & \mathcal{L}_{i}.\theta = \mathcal{L}_{i-1}.\theta + f(\mathcal{R}_i,\mathcal{L}_{i-1})*\arcsin(\frac{d(P_{s+i}, \mathcal{L}_{i-1})}{j*\zeta/2})/j \\	
            \end{aligned}
        \right]\hspace{0ex}~else\hspace{1ex}(3)\\
    \end{aligned}
    \right.
\end{equation*}
\vspace{-2ex}
\end{small}

\ni where (a) $1 \le i \le k+1$; (b) $\mathcal{R}_{i-1}$ = $\vv{P_sP_{s+i-1}}$, is the directed line segment whose end point $P_{s+i-1}$ is in $\dddot{\mathcal{T}_s}[P_s, \ldots, P_{s+k}]$; (c) $\mathcal{L}_{i}$ is the directed line segment built by fitting function $\mathbb{F}$ to fit sub-trajectory $\dddot{\mathcal{T}_s}[P_s, \ldots, P_{s+i}]$ and $\mathcal{L}_{0}$ = $\mathcal{R}_{0}$; (d) $j = \lceil(|\mathcal{R}_{i}|*2/\zeta - 0.5)\rceil$; (e) $f()$ is a sign function such that $ f(\mathcal{R}_i,\mathcal{L}_{i-1})$ = $1$ if the included angle $\angle(\mathcal{R}_{i-1}, \mathcal{R}_{i})$ = $(\mathcal{R}_i.\theta - \mathcal{L}_{i-1}.\theta)$ falls in the range of $(-2\pi, -\frac{3\pi}{2}]$, $[-\pi, -\frac{\pi}{2}]$, $[0, \frac{\pi}{2}]$ and $[\pi, \frac{3\pi}{2})$, and $f(\mathcal{R}_i,\mathcal{L}_{i-1})$ = $-1$, otherwise; (f) $\zeta/2$ is a step length to control the increment of $|\mathcal{L}|$.

\begin{figure}[tb!]
\vspace{-0.5ex}
\centering
\includegraphics[scale = 0.6]{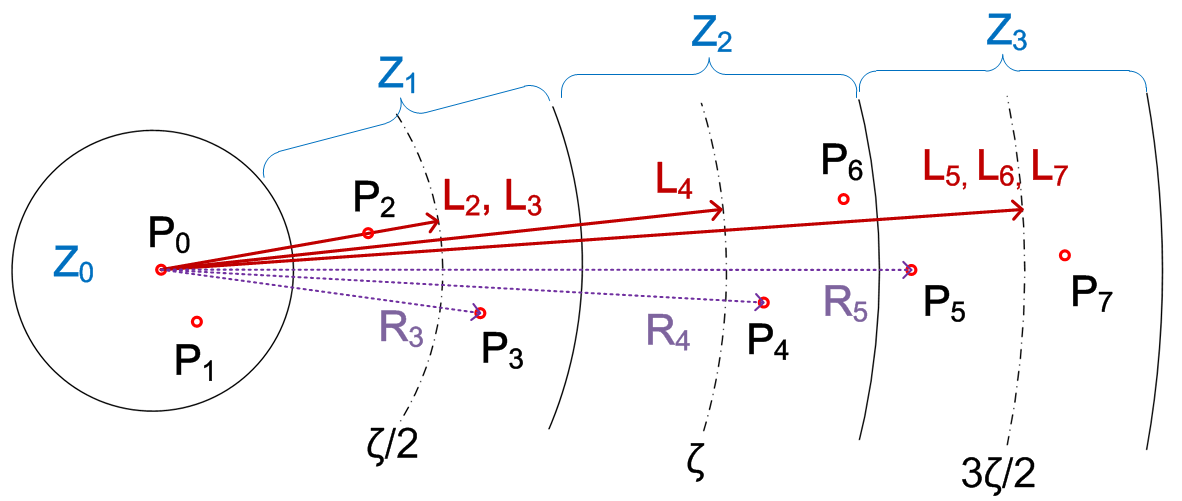}
\vspace{-2.5ex}
\caption{\small An example of the fitting function}
\label{fig:fitfunction}
\vspace{-3ex}
\end{figure}

Given any sub-trajectory $\dddot{\mathcal{T}_s}[P_s$, $\ldots$, $P_{s+k}]$ and any error bound $\zeta$, the expression $j = \lceil(|\mathcal{R}_{i}|*2/\zeta - 0.5)\rceil$ in the fitting function $\mathbb{F}$ essentially partitions the space into zones around the center point $P_s$ such that for each $j\ge 0$, zone $Z_j$ = $\{P_j \ |\ j*\zeta/2-\zeta/4 < |\vv{P_sP_j}|\le j*\zeta/2+\zeta/4\}$, \ie the radii of $Z_0, Z_1, Z_2$ and $Z_3$ to $P_s$ are in the ranges of $(-\frac{1}{4}\zeta, \frac{1}{4}\zeta]$, $(\frac{1}{4}\zeta, \frac{3}{4}\zeta]$, $(\frac{3}{4}\zeta, \frac{5}{4}\zeta]$ and $(\frac{5}{4}\zeta, \frac{7}{4}\zeta]$, respectively, and all ranges have a fixed size $\zeta/2$ as shown in Figure~\ref{fig:fitfunction}. Moreover, the angle of the  directed line segment $\mathcal{L}_i$ is adjusted from $\mathcal{L}_{i-1}$ to make it closer to $P_{s+i}$ than $\mathcal{L}_{i-1}$, \ie $d(P_{s+i}, \mathcal{L}_{i}) \le d(P_{s+i}, \mathcal{L}_{i-1})$ for any $i \in (s, s+k]$, and the angle from $\mathcal{L}_{1}$ to $\mathcal{L}_{k}$ is bounded by a constant (Lemma~\ref{lemma-stepwise-angle}).


The fitting function $\mathbb{F}$ also creates a virtual stepwise sub-trajectory $\dddot{\mathcal{T}_v}[V_s$, $\ldots$, $V_{s+l}]$ such that $V_s = P_s$, $|\vv{V_sV_{s+j}}|$ = $\zeta/2*j$ $(j\in[0,l], 0<l\le k)$.
For each point $P_{s+i}$ in the sub trajectory, it is mapped to a virtual point $V_{s+j}$ in $\dddot{\mathcal{T}_v}$ locating in zone $Z_j$.
Observe that (a) it is possible that $(|\mathcal{R}_{i}| - |\mathcal{L}_{i-1}|) \le 0$, (b) the fitting function  $\mathbb{F}$ forms a directed line segment $\mathcal{L}_i$, which is closer to $\mathcal{R}_i$ than $\mathcal{L}_{i-1}$, and partitions the points in the sub-trajectory $\dddot{\mathcal{T}}$ into two classes:

\stab (1) {\em Active points}. Points $P_{s}$ and $P_{s+i}$ such that $|\mathcal{R}_{i}| - |\mathcal{L}_{i-1}|$ $>$ $\zeta/4$ are referred to as active points. An active point $P_{s+i}$ is mapped to the virtual point in zone $Z_j$ with $j= \lceil(|\mathcal{R}_{i}|*2/\zeta - 0.5)\rceil$, and each zone has at most one active point.

\stab(2) {\em Inactive points}. Points $P_{s+i}$ such that $|\mathcal{R}_{i}| - |\mathcal{L}_{i-1}| \le \zeta/4$ are referred to as inactivated points. For an inactive point $P_{s+i}$, it is mapped to zone $Z_j$ with $j = |\mathcal{L}_{i-1}|*2/\zeta$. There may exist none or multiple inactive points in a zone.

We next explain the fitting function $\mathbb{F}$ with an example.

\begin{example}
\label{exm-fittingfunction}
Consider the sub-trajectory $[P_0,\ldots, P_7]$  in Figure~\ref{fig:fitfunction} whose eight points fall in zones $Z_0, Z_1, Z_2, Z_3$.

\sstab (1) Point $P_0$ is the start point and the first active point, and $\mathcal{L}_0 = \mathcal{R}_0 = \vv{P_0P_0}$.

\sstab (2) Point $P_1$ is inactive in zone $Z_0$, as $|\mathcal{R}_{1}|$ = $|\vv{P_0P_1}|$ $<$ $\frac{\zeta}{4}$ and $(|\mathcal{R}_{1}| - |\mathcal{L}_{0}|) \le \frac{\zeta}{4}$. Hence, $\mathcal{L}_{1} = \mathcal{L}_{0}$  (case (1)).

\sstab (3) Point $P_2$ is active in $Z_1$, as $|\mathcal{R}_{2}| >  \frac{\zeta}{4}$ and $|\mathcal{L}_{1}|=0$ .
Hence, $|\mathcal{L}_{2}|  = \frac{\zeta}{2}$ and $\mathcal{L}_{2}.\theta = \mathcal{R}_{2}.\theta$ (case 2).

\sstab (4) Point $P_3$ is inactive in $Z_1$, as $(|\mathcal{R}_{3}| - |\mathcal{L}_{2}|) \le \frac{\zeta}{4}$. Hence, $\mathcal{L}_{3} = \mathcal{L}_{2}$  (case 1).

\sstab (5) Point $P_4$ is active in $Z_2$, as $|\mathcal{R}_{4}| - |\mathcal{L}_{3}|) > \frac{\zeta}{4}$ and $|\mathcal{L}_{3}|\ne 0$. Hence, $|\mathcal{L}_{4}|  = 2*\frac{\zeta}{2} = \zeta$ and the angle of $\mathcal{L}_4$ is also calculated accordingly (case 3).

\sstab (6) Similarly, point $P_5$ is active in $Z_3$ (case 3), and points $P_6$ and $P_7$ are inactive  (case 1). Here $P_6$ is mapped to $Z_3$ as $|\mathcal{L}_{5}| = \frac{3}{2}\zeta$ though it is physically located in zone $Z_2$.
\end{example}
\vspace{-1ex}

\subsection{Analyses of the Fitting Function}
\label{subsec-fittingfunction-analysis}
We next give an analysis of the fitting function $\mathbb{F}$.
 First, by the definition of $\mathbb{F}$, it is easy to have the following.

\begin{pprop}
\label{prop-fitting-func-time}
Given any sub-trajectory $\dddot{\mathcal{T}_s}[P_s, \ldots, P_{s+k}]$ and error bound $\zeta$, the directed line segment $\mathcal{L}_{i}$ $(i\in[1,k])$ can be computed by the fitting function $\mathbb{F}$  in $O(1)$ time.
\end{pprop}

The fitting function $\mathbb{F}$ also enables a local distance checking method, as shown below.

\begin{ttheorem}
\label{prop-fitting-function}
Given any sub-trajectory $\dddot{\mathcal{T}_s}[P_s, \ldots, P_{s+k}]$ with $k\le 4\times 10^5$ and error bound $\zeta$,
then $d(P_{s+i}, {\vv{P_{s}P_{s+k}})}\le \zeta$ for each $i\in[0, k]$ if $P_{s+k}$ is an active point and $d(P_{s+i}, \mathcal{L}_{i-1})\le \zeta/2$ for each $i\in[1, k]$.
\end{ttheorem}

To prove Theorem~\ref{prop-fitting-function}, we first introduce a special class of trajectories, based on which we show that Theorem~\ref{prop-fitting-function} holds.

\stitle{Stepwise trajectories}. We say that a trajectory $\dddot{\mathcal{T}}[P_s$, $\ldots$, $P_{s+k}]$ is stepwise \wrt $\zeta/2$ if and only if $|\mathcal{R}_{i}| = i*\zeta/2$ for each directed line segment $\mathcal{R}_{i}$ = $\vv{P_sP_{s+i}}$ ($i\in[0,k]$).

Observe that $|\mathcal{R}_{i}|$  $-$ $|\mathcal{R}_{i-1}|$ = $\zeta/2$  and $\lceil|\mathcal{R}_{i}|*2/\zeta - 0.5\rceil = i$ ($i\in[1, k]$).
Hence, for stepwise sub-trajectories  $\dddot{\mathcal{T}}[P_s$, $\ldots$, $P_{s+k}]$, the  fitting function can be simplified as  $\mathbb{F}'$ below.

\begin{small}
\vspace{-1ex}
\begin{equation*}
\hspace{-1.5ex} \left\{
    \begin{aligned}
        &\hspace{-1.5ex}\left[
            \begin{aligned}
            & \mathcal{L}_{1} = {\zeta}/{2}\\
            & \mathcal{L}_{1}.\theta = \mathcal{R}_{1}.\theta\\
             \end{aligned}
        \right]\hspace{46.5ex}~i=1\hspace{1ex}(1)\\
        &\hspace{-1.5ex}\left[
            \begin{aligned}
            & |\mathcal{L}_{i}|  = i*{\zeta}/{2}\\
            & \mathcal{L}_{i}.\theta = \mathcal{L}_{i-1}.\theta + f(\mathcal{R}_i,\mathcal{L}_{i-1})*\arcsin(\frac{d(P_{s+i}, \mathcal{L}_{i-1})}{i*\zeta/2})/i \\	
            \end{aligned}
        \right]\hspace{.5ex}i\ge 2\hspace{1ex}(2)~ \\
    \end{aligned}
    \right.
\end{equation*}
\vspace{-1ex}
\end{small}

Stepwise trajectories have the following properties.

\begin{llemma}
\label{lemma-stepwise-angle}
Given any sub-trajectory $\dddot{\mathcal{T}_s}[P_s, \ldots, P_{s+k}]$ and error bound $\zeta$,
if $d(P_{s+i}, \mathcal{L}_{i-1})\le \frac{\zeta}{2}$  for each $i$ $\in[2, k]$, then the angle change between $\mathcal{L}_{1}$ and $\mathcal{L}_{k}$ is bounded by $\Delta\theta$ = $\lim\limits_{k \to \infty }{\sum_{i=2}^k\frac{1}{i} * {\arcsin(\frac{1}{i})}} < 0.8123$ (or $46.54^o$).
\end{llemma}

\begin{proof}
By the revised fitting function $\mathbb{F'}$ for a stepwise sub-trajectory and $d(P_{s+i}, \mathcal{L}_{i-1})\le \frac{\zeta}{2}$ for all $i$ $\in[2, k]$,
we have $\Delta\theta$ $\le {\sum_{i=2}^{k}\frac{1}{i} * {\arcsin(\frac{1}{i})}}$, which is  monotonically increasing with the increment of $k$. 
As $x\le \arcsin(x) \le {x}/{\sqrt{1-x^2}}$ ($0\le x< 1$), we also have $\frac{1}{i} \le \arcsin\frac{1}{i}\le$ ${1}/{\sqrt{i^2-1}}$ ($i\ge 2$).

%
%
Hence, we have $\Delta\theta \le \lim\limits_{k \to \infty }{\sum_{i=2}^{k}(\frac{1}{i}*\frac{1}{\sqrt{i^2-1}})}$
$< {\int_{2}^{\infty}(\frac{1}{x}*\frac{1}{\sqrt{x^2-1}})}dx+\frac{1}{2}*\frac{1}{\sqrt{2^2-1}}=\frac{\pi}{6}+\frac{1}{2\sqrt{3}} \approx 0.8123$.
\end{proof}


\begin{figure}[tb!]
\centering
\includegraphics[scale = 0.68]{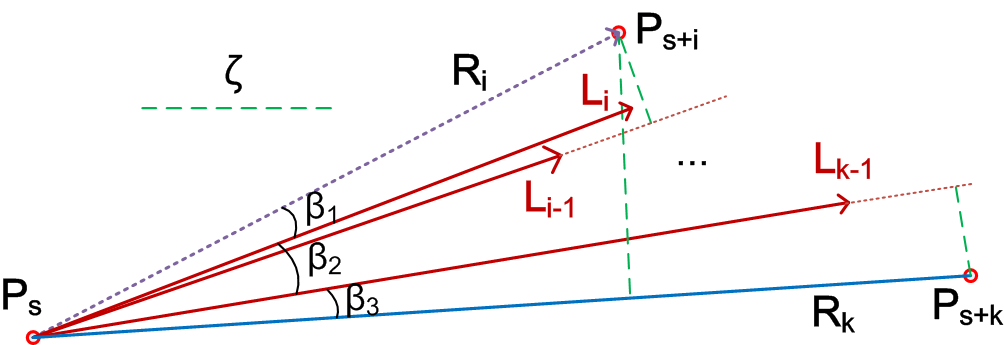}
\vspace{-2ex}
\caption{\small Example for the proof of Lemma~\ref{lemma-stepwise-errorbound}}
\label{fig:stepwise-maxdistance}
\vspace{-4ex}
\end{figure}

\vspace{-1ex}
\begin{llemma}
\label{lemma-stepwise-errorbound}
Given any sub-trajectory $\dddot{\mathcal{T}_s}[P_s, \ldots, P_{s+k}]$ with $k\le 4\times 10^5$ and error bound $\zeta$,
then $d(P_{s+i}, {\vv{P_{s}P_{s+k}})}\le \zeta$ for each $i\in[0, k]$ if $d(P_{s+i}, \mathcal{L}_{i-1})\le \frac{\zeta}{2}$ for each $i\in[1, k]$.
\end{llemma}

\begin{proof}
Consider the four directed line segments  $\mathcal{R}_{i} = \vv{P_sP_{s+i}}$, $\mathcal{R}_{k} = \vv{P_sP_{s+k}}$,  $\mathcal{L}_{i-1}$ and  $\mathcal{L}_{k-1}$ shown in Figure~\ref{fig:stepwise-maxdistance}.
Further, let $\beta_1$ = $\mathcal{R}_{i}.\theta - \mathcal{L}_{i-1}.\theta$, $\beta_2$ = $\mathcal{L}_{k-1}.\theta - \mathcal{L}_{i}.\theta$, $\beta_3$ = $\mathcal{R}_{k}.\theta - \mathcal{L}_{k-1}.\theta$.
We then adjust the included angles $\beta_1$ as follows:
(a) if $\frac{\pi}{2}<|\beta_1|\le \pi$,  $\beta_1$ = $\pi$ - $\beta_1$,
(b) if $\pi<|\beta_1|\le \frac{3}{2}\pi$,  $\beta_1$ = $\beta_1$ - $\pi$,
(c) if $\frac{3}{2}\pi<|\beta_1|\le {2}\pi$,  $\beta_1$ = $2\pi$ - $\beta_1$, and
(d) $\beta_1$ = $|\beta_1|$, otherwise.
 The included angle $\beta_3$ is adjusted along the same line as $\beta_1$, and $\beta_2$ is bounded by Lemma~\ref{lemma-stepwise-angle}.

Observe that $d(P_{s+i}, \mathcal{R}_{k})\le i*\frac{\zeta}{2}*\sin(|\beta_1| + |\beta_2| + |\beta_3|)$
$\le$ $i*\frac{\zeta}{2}*\sin(\arcsin\frac{1}{i}+\sum_{j=i+1}^{k-1}{(\frac{1}{j}*\arcsin\frac{1}{j})}+\arcsin\frac{1}{k})$.
For any $k\le 4\times 10^5$, $i*\sin(|\beta_1| + |\beta_2| + |\beta_3|)<2$, and, hence,
we have $d(P_{s+i}, \mathcal{R}_{k})< \frac{\zeta}{2}*2 = \zeta$.
\end{proof}

By mapping inactive and active points of a trajectory to virtual points of a trajectory stepwise \wrt ${\zeta}/2$, one can readily prove Theorem~\ref{prop-fitting-function} along the lines as Lemma~\ref{lemma-stepwise-errorbound}.

\stitle{Remarks}.
(1) Our fitting function achieves local distance checking, as indicated by Proposition~\ref{prop-fitting-func-time} and Theorem~\ref{prop-fitting-function};
(2) For a sub-trajectory $\dddot{\mathcal{T}_s}[P_s, \ldots, P_{s+k-1}]$ represented by a single directed line segment,
 we restrict $k\le 4*10^5$, which suffices  for the need of trajectory simplification in practice.

\subsection{Algorithm OPERB}
\label{subsec-algorithm}

We are now ready to present our one-pass error bounded algorithm, which makes use of the local distance checking method based on the fitting function $\mathbb{F}$.

The main result of this section is as follows.

\begin{ttheorem}
\label{thm-operb}
Given any trajectory $\dddot{\mathcal{T}}[P_0, \ldots, P_{n}]$ and error bound $\zeta$, there exists a one-pass trajectory simplification algorithm that is error bounded by $\zeta$.
\end{ttheorem}

We prove Theorem~\ref{thm-operb} by providing such an algorithm for  trajectory simplification, referred to as \operb shown in Figure~\ref{alg:operb}. Given
a trajectory $\dddot{\mathcal{T}}[P_0, \ldots, P_n]$ and an error bound $\zeta$ as input, algorithm \operb outputs a compression trajectory, \ie a piecewise line representation $\overline{\mathcal{T}}$ of $\dddot{\mathcal{T}}$.

We first describe its procedure, and then present \operb.

\stitle{Procedure \kw{getActivePoint}}. It takes as input a trajectory $\dddot{\mathcal{T}}$, a start point $P_s$, the current active point $P_a$, the current directed line segment $\mathcal{L}_a$ and the error bound $\zeta$, and finds the next active point $P_i$. (1) When $P_i = \kw{nil}$, it means that no more active points could be found in the remaining sub-trajectory $\dddot{\mathcal{T}}[P_s,\ldots,P_n]$; (2) When $flag = \True$, it means that the next active point $P_i$ can be combined with the current directed line segment $\mathcal{L}_a$ to form a new directed line segment; Otherwise, (3) a new line segment should be generated, and a new start point is considered.

It first increases $a$ by $1$ as it considers the data points after $P_a$, and sets $flag$ to \True\ (line 1).
Secondly, by the definition of the fitting function $\mathbb{F}$, it finds the next active point $P_i$ (lines 2--{6}).
\eat{More specifically, if $(|\mathcal{R}_i| - |\mathcal{L}_a|) \le \zeta/4$ and $d(P_i, \mathcal{L}_a) > \zeta/2$ (or $d(P_i, \mathcal{R}_a) > \zeta$), then $flag$ is set to \False~(lines 3--4); if $(|\mathcal{R}_i| - |\mathcal{L}_a|) > \zeta/4$ and $d(P_i, \mathcal{L}_a) > \zeta/2$, then $flag$ is also set to \False~(lines 6).}
Thirdly, if $i = n + 1$, then all data points in $\dddot{\mathcal{T}}$ have been considered, hence, $P_i$ is set to \kw{nil}(line 7).
Finally, $(P_i, flag)$ is returned (line 8).

\stitle{Algorithm \operb}.  It takes as input a trajectory $\dddot{\mathcal{T}}$ and an error bound $\zeta$, and returns the simplified trajectory $\overline{\mathcal{T}}$.

After initializing (line 1), it then repeatedly processes the data points in $\dddot{\mathcal{T}}$ one by one until all data points have been considered, \ie $P_a = \kw{nil}$ (lines 2--8).
\eat{The start point $P_s$ of a new directed line segment is always set to the last active point $P_e$, and directed line segment $\mathcal{L}_a$ is formed by the fitting function $\mathbb{F}$ (line 3). Note that here $\mathcal{L}_{a}$ = $\mathbb{F}(P_a, \vv{P_sP_s})$ is an abbreviation, which needs to recursively call $\mathbb{F}$ to compute $\mathcal{L}_{a}$. It next calls \kw{getActivePoint} to get the next active point $P_a$. }
If $P_a$ is not $\kw{nil}$ and $flag$ is \True, it means that $P_a$ can be combined with the current directed line segment $L_a$ (lines 5--7).
If $flag$ is \False, then a directed line segment $\vv{P_sP_e}$ is generated and added to $\overline{\mathcal{T}}$ (line 8).
Finally, the set $\overline{\mathcal{T}}$ of directed line segments, \ie a piecewise line segmentation of $\dddot{\mathcal{T}}$, is returned (line 9).

\begin{figure}[tb!]
\begin{center}
{\small
\begin{minipage}{3.36in}
\myhrule \vspace{-2ex}
\mat{0ex}{
{\bf Algorithm}~$\operb(\dddot{\mathcal{T}}[P_0,\ldots,P_n], \zeta)$\\
\sstab
\bcc \hspace{1ex}\=$\overline{\mathcal{T}}$ := $\emptyset$; $P_e := P_0$; $(P_a,flag)$ := $\kw{getActivePoint}(\dddot{\mathcal{T}}, P_0, P_0, \mathcal{L}_{0}, \zeta)$;\\
\icc \>$\While P_a \ne \kw{nil}\ \Do$ \{\\
\icc \>\hspace{4ex}\=$P_s := P_e$; \ \ $\mathcal{L}_{a}$ = $\mathbb{F}(P_a, \vv{P_sP_s})$;\\
\icc \>\>$(P_a,flag)$ := $\kw{getActivePoint}(\dddot{\mathcal{T}}, P_s, P_a, \mathcal{L}_{a}, \zeta)$;\\
\icc \>\>\While $P_a \ne \kw{nil} ~\&~ flag=\True$ \Do \{\\
\icc \>\>\hspace{4ex}\=$\mathcal{L}_{a}$ := $\mathbb{F}(P_a, \mathcal{L}_{a})$;\ \   $P_e := P_a$;  \\
\icc \>\>\>$(P_a,flag)$ := $\kw{getActivePoint}(\dddot{\mathcal{T}}, P_s, P_a, \mathcal{L}_{a}, \zeta)$;\ \}\\
\icc \>\>$\overline{\mathcal{T}}$ := $\overline{\mathcal{T}}\cup\{\vv{P_sP_e}\}$;\ \}\\
\icc \>\Return $\overline{\mathcal{T}}$.\\
\sstab
{\bf Procedure}~$\kw{getActivePoint}(\dddot{\mathcal{T}}, P_s, P_a, \mathcal{L}_a, {\zeta})$\\
\sstab
\bcc \hspace{1ex}\=$i := a+1$;\ \ $flag$ := \True; \\
\icc\> \While($(|\mathcal{R}_i| - |\mathcal{L}_a|) \le \zeta/4$ \& $i\le n$ \&  $(i-s)\le 4\times 10^5$ \Do\{\\
\icc \>\hspace{4ex}\=\If $d(P_i, \mathcal{L}_a) > \zeta/2$ \Or $d(P_i, \mathcal{R}_a) > \zeta$ \Then \\
\icc \>\>\hspace{4ex}\=$flag$ := \False; \ \ \Break; \\
\icc \>\>$i$ := $i + 1$; \}\\
\icc \>\If $d(P_i, \mathcal{L}_a) > \zeta/2$   \&  $|\mathcal{L}_a|>0$  \Then $flag$ := \False;\\
\icc \>\If $i = n + 1$  \Then $P_i$ :=\kw{nil};\\
\icc \>\Return $(P_i, flag)$.
}
\vspace{-2.5ex}
\myhrule
\end{minipage}
}
\end{center}
\vspace{-3.5ex}
\caption{\small Algorithm \operb}\label{alg:operb}
\vspace{-3ex}
\end{figure}

We next explain algorithm \operb with an example.

\begin{figure}[tb!]
\centering
\includegraphics[scale = 0.64]{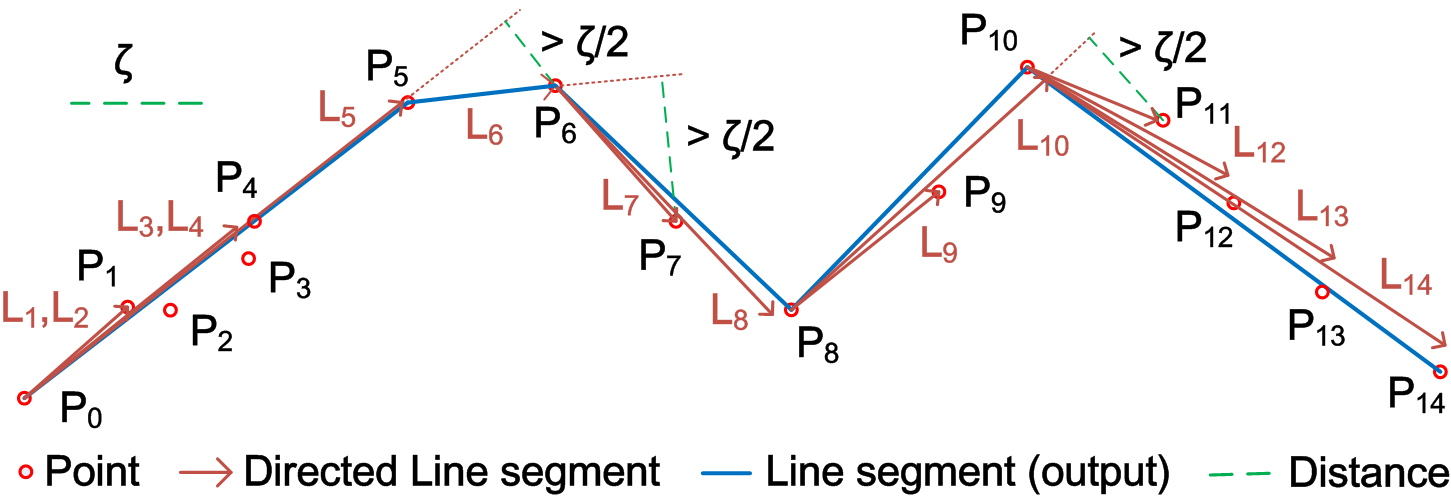}
\vspace{-1.5ex}
\caption{\small A running example of algorithm \operb.}\label{exa-fig:operb}
\vspace{-3.5ex}
\end{figure}

\begin{example}
\label{exam-alg-operb}
Algorithm  \operb takes as input the trajectory and $\zeta$ shown in Figure~\ref{fig:dp}, and its output is illustrated in Figure~\ref{exa-fig:operb}.
The process of algorithm  \operb is as follows.

\sstab(1) It initializes $\overline{\mathcal{T}}$  with $\emptyset$, the last active point $P_e$ with $P_0$ and the current active point $P_a$ with $P_1$ (line 1).

\sstab(2) As $P_a \ne \kw{nil}$ (line 2), it then sets $P_s = P_e = P_0$ and $\mathcal{L}_a = \mathbb{F}(P_a, \vv{P_sP_s}) =\mathbb{F}(P_1, \vv{P_0P_0})$ = $\mathcal{L}_1$ (line 3).

\sstab(3) It then calls $\kw{getActivePoint()}$ to get the next active point $P_a = P_3$ and $flag = \True$ (line 4). As $flag = \True$ means that $P_3$ can be combined with the current line segment $\mathcal{L}_a$ = $\mathcal{L}_1$, so it updates $\mathcal{L}_a$  to $\mathcal{L}_3$, and $P_e$ to $P_3$ (line 6).

\sstab(4) Then it continues reading the next active point $P_a$ = $P_5$ with $flag = \True$  (line 7), and updates the current line segment $\mathcal{L}_a$ to $\mathcal{L}_5$, and $P_e$  to $P_5$ (lines 5, 6).

\sstab(5) It gets the next active $P_a = P_6$ and $flag = \False$,  as $d(P_6, \mathcal{L}_5)>\frac{\zeta}{2}$, meaning that $P_6$ should not be compressed to the current directed line segment (line 5). Hence, it adds $\vv{P_sP_e} = \vv{P_0P_5}$ to $\overline{\mathcal{T}}$ (line 8), sets $P_s = P_e = P_5$, and updates $\mathcal{L}_a$ to $\mathbb{F}(P_a, \vv{P_sP_s}) =  \mathbb{F}(P_6, \vv{P_5P_5})$ = $\mathcal{L}_6$ (line 3).

\sstab(6) The process continues until all points have been processed. At last, the algorithm outputs five continuous line segments $\{\vv{P_0P_5}$, $\vv{P_5P_6}$, $\vv{P_6P_8}$, $\vv{P_8P_{10}}$, $\vv{P_{10}P_{14}}\}$.
\end{example}

\stitle{Correctness \& complexity analysis}. The correctness of algorithm \operb follows from  Theorem~\ref{prop-fitting-function} immediately.
 Observe that for a trajectory $\dddot{\mathcal{T}}$ with $n$ data points, the fitting function $\mathbb{F}$ is called at most $n$ times, and each data point is considered once and only once. By Proposition~\ref{prop-fitting-func-time}, algorithm \operb runs in $O(n)$ time.
It is also easy to verify that algorithm \operb takes $O(1)$ space, as the directed line segment in $\overline{\mathcal{T}}$ can be output immediately once it is generated.

Note that this also completes the proof of Theorem~\ref{thm-operb}.

\subsection{Optimization Techniques}
\label{subsec-optimizations}

We further propose five optimization techniques for \operb to achieve a better compression ratio. The key idea behind this is to (1) compress as many points as possible with a directed line segment, or (2) to let the directed line segment $\mathcal{L}_i$ as close as possible to the current active point $P_i$ so that it has a higher possibility to represent $P_{i+1}$. These optimization techniques are organized by the processing order from the start to the end points of directed line segments.

\stitle{(1) Choosing the first active point after $P_s$}. Algorithm~\operb calls procedure $\kw{getActivePoint}$ to get the first active point $P_a$ in a sub-trajectory $\dddot{\mathcal{T}}[P_{s+1}, \ldots,$ $P_{s+k}]$ such that $|\vv{P_sP_{a}}|>\zeta/4$ (line 4 in Figure~\ref{alg:operb}).
However, as indicated by the fitting function $\mathbb{F}$, we can replace $P_a$ with the first point $P_{b}$ such that $|\vv{P_sP_{b}}|>\zeta$, without affecting the boundness of algorithm~\operb.
This method potentially improves the compression ratio because more points are covered by the directed line segment $\vv{P_sP_b}$ than $\vv{P_sP_a}$,
and $\vv{P_sP_b}$ is also closer to $P_b$ than $\vv{P_sP_a}$.


\stitle{(2) Adjusting the distance condition}. Given any sub trajectory $\dddot{\mathcal{T}}[P_s, \ldots, P_{s+k}]$ and error bound $\zeta$, let $d^+_{max}$ = $\max\{d(P_{s+i}, \mathcal{L}_{i-1}) ~|~ f(\mathcal{R}_i,\mathcal{L}_{i-1})=1 ~and~ i \in [1, k]\}$ and $d^-_{max}$ = $\max \{d(P_{s+i}, \mathcal{L}_{i-1}) ~|~f(\mathcal{R}_i,\mathcal{L}_{i-1})=-1 ~and~ i \in [s+1, s+k]\}$. Then the condition $d(P_{s+i}, \mathcal{L}_{i-1}) \le \zeta/2$ in Theorem~\ref{prop-fitting-function} can be replaced with $d^-_{max} + d^+_{max} \le \zeta$, and algorithm \operb remains error bounded by $\zeta$.
Say, if $d^+_{max} = 0.3\zeta$ and  $d^-_{max} = 0.6\zeta$, then $d(P_{s+i}, \mathcal{L}_{k})$ for each $i \in [s, s+k]$ is still less than $0.3\zeta + 0.6\zeta= 0.9\zeta < \zeta$.

Note that  $d(P_{s+i}, \mathcal{L}_{i-1}) \le \zeta/2$ implies $d^+_{max} \le \zeta/2$ and $d^-_{max} \le \zeta/2$, and, hence, $d^-_{max} + d^+_{max} \le \zeta$.
Therefore, $d(P_{s+i}, \mathcal{L}_{i-1}) \le \zeta/2$ is a special case of  $d^-_{max} + d^+_{max} \le \zeta$.

\stitle{(3) Making $\mathcal{L}$ more close to the active points}. When \operb calculates the angle $\mathcal{L}_{i}.\theta$ of an active point $P_i$, the factor $d(P_i, \mathcal{L}_{i-1})$ in the fitting function $\mathbb{F}$ can be replaced by a bigger number $d_x$ such that $0 \le d_x \le d^-_{max}$ when $f(\mathcal{R}_i,\mathcal{L}_{i-1})=-1$ or $0 \le d_x \le d^+_{max}$ when $f(\mathcal{R}_i,\mathcal{L}_{i-1})=1$, to let $\mathcal{L}_i$ be more close to $P_i$, under the restriction that $(\arcsin(\frac{d_x}{j*\zeta/2})/j)$ is not larger than $\arcsin(\frac{d(P_i, \mathcal{L}_{i-1})}{j*\zeta/2})$.

\stitle{(4) Incorporating missing active points}. For a sub-trajectory $\dddot{\mathcal{T}}[P_s$, $\ldots, P_a, \ldots, P_{a+i},\ldots, P_{s+k}]$, whereas $P_{a}$ and $P_{a+i}$ are two consecutive active points. Let $j_a$ = $\lceil(|\mathcal{R}_{a}|*2/\zeta - 0.5)\rceil$, $j_{a+1}$ = $\lceil(|\mathcal{R}_{a+i}|*2/\zeta - 0.5)\rceil$ and $\Delta j = j_{a+1} - j_a$. If $\Delta j > 1 $, then there are no active points between zones $Z_{j_a}$ and $Z_{j_{a+1}}$. In this case, we replace $\mathcal{L}_{a+i}.\theta$ with $\mathcal{L}_{a+i-1}.\theta + f(\mathcal{R}_{a+i},\mathcal{L}_{a+i-1})*\arcsin(\frac{d(P_{a+i}, \mathcal{L}_{a+i-1})}{j_{a+1}*\zeta/2})*\frac{\Delta j}{j_{a+1}}$ for the fitting function $\mathbb{F}$, instead of $\mathcal{L}_{a+i-1}.\theta + f(\mathcal{R}_{a+i},\mathcal{L}_{a+i-1})*\arcsin(\frac{d(P_{a+i}, \mathcal{L}_{a+i-1})}{j_{a+1}*\zeta/2})*\frac{1}{j_{a+1}}$ to compensate the side effects of missing active points to make the line $\mathcal{L}_{a+i}$ more closer to $P_{a+i}$. Note that $\mathcal{L}_{a+i-1} = \mathcal{L}_{a}$ and $\Delta j>0$. Moreover, $d(P_{a+i}, \mathcal{L}_{a+i-1})$ could also be replaced by $d_x$ as above.

\stitle{(5) Absorbing data points after $P_{s+k}$}.
Given any sub-trajectory $\dddot{\mathcal{T}_s}[P_s, \ldots, P_{s+k},\ldots,P_{s+t}]$ and error bound $\zeta$, if $[P_s, \ldots, P_{s+k}]$ is compressed to a line segment $\vv{P_{s}P_{s+k}}$, then any point $P_{s+t}$ ($t>k$) can also be compressed to $\vv{P_{s}P_{s+k}}$ as long as $d(P_{s+t}, {\vv{P_{s}P_{s+k}}})\le\zeta$ such that \operb can compress more points into the directed line segment.


\stitle{Remark}. These optimization techniques are seamlessly integrated into \operb, and Theorem~\ref{thm-operb} remains intact.


\vspace{-0.5ex}
\section{An Aggressive Approach}
\label{sec-operba-agg}
In this section, we introduce an aggressive one-pass trajectory simplification algorithm, referred to as \operba, which extends algorithm \operb by further allowing trajectory interpolation under certain conditions, and even achieves a better compression ratio than algorithm \dpa, the existing \lsa algorithm with the best compression ratio.


\subsection{Trajectory Interpolation}
\label {subsec-interpolation}

Existing line simplification algorithms, even the global distance checking algorithm \dpa and our algorithm \operb, may generate a set of {\em anomalous line segments}.

\stitle{Anomalous line segments}. Consider a trajectory $\dddot{\mathcal{T}}$$[P_0,$ $\ldots, P_n]$ and its piece-wise line representation $\overline{\mathcal{T}}[\mathcal{R}_0,$ $\ldots , \mathcal{R}_m]$ ($0< m \le n$) generated by an \lsa algorithm. A line segment $\mathcal{R}_i$ ($i\in[0,m]$) is anomalous if it only represents two data points in $\dddot{\mathcal{T}}$, \ie its own start and end points.

Anomalous line segments impair the effectiveness of trajectory simplification. 
We illustrate this with an example.

\begin{figure}[tb!]
\centering
\includegraphics[scale = 0.64]{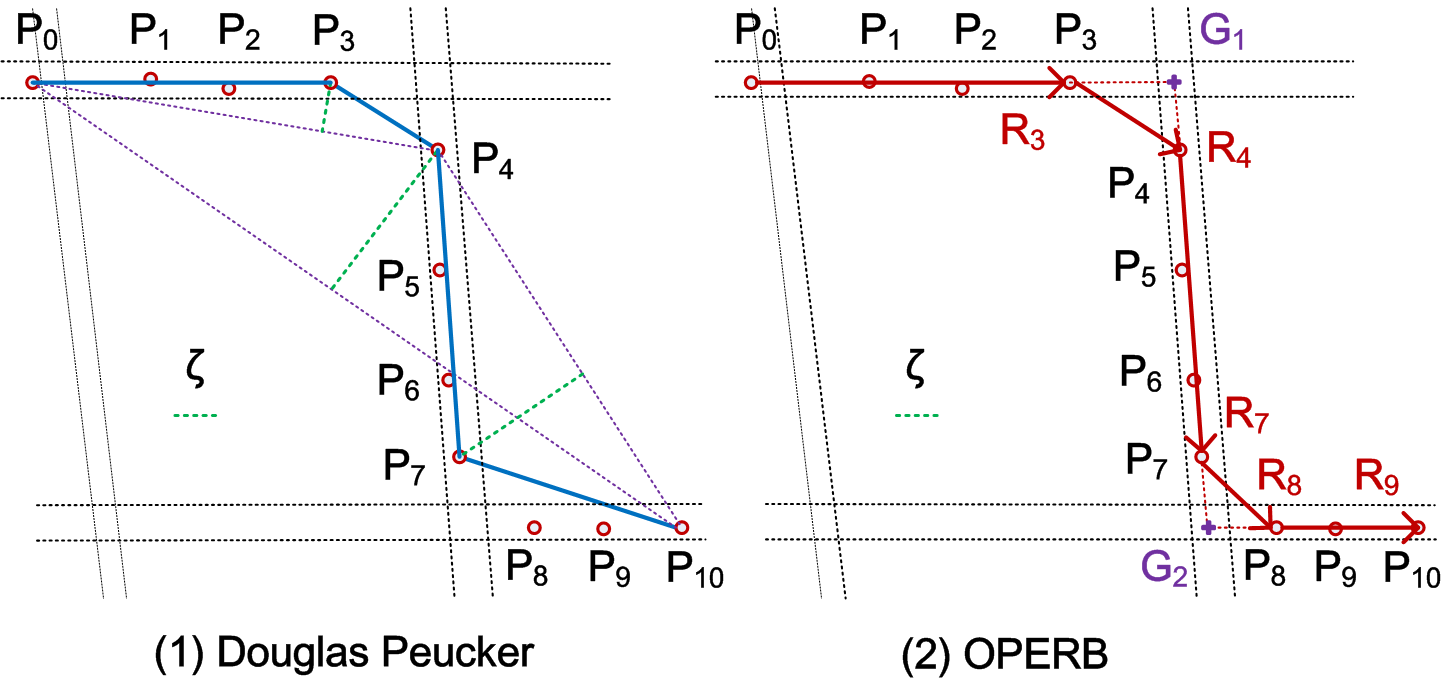}
\vspace{-2ex}
\caption{\small Example anomalous line segments: the compression results of algorithms \dpa and \operb on a sub-trajectory with eleven points collected from a moving vehicle running on an urban road network.}
\label{fig:onroad}
\vspace{-3.5ex}
\end{figure}

\begin{example}
\label{exm-anomalous-ls}
Let us consider the compressing results of algorithms \dpa and \operb on a sub-trajectory, shown in Figure~\ref{fig:onroad}, which has one crossroad between data points $P_3$ and $P_4$, and another crossroad between data points $P_7$ and $P_8$.
Given the sub-trajectory and the error bound as shown in Figure~\ref{fig:onroad}, algorithm \dpa returns four directed line segments $\{\vv{P_0P_3}$, $\vv{P_3P_4}$, $\vv{P_4P_7}$, $\vv{P_7P_{10}}\}$, and algorithm \operb outputs five directed line segments $\{\vv{P_0P_3}$, $\vv{P_3P_4}$, $\vv{P_4P_7}$, $\vv{P_7P_8}$, $\vv{P_8P_{10}}\}$, respectively.
Observe that the directed line segments $\vv{P_3P_4}$ and $\vv{P_7P_8}$ are anomalous. 
\end{example}

In this work, we propose the use of interpolating new data points, referred to as {\em patch points}, into a trajectory under certain conditions to reduce the number of anomalous line segments to a large extent.
The rational behind this is that moving objects have sudden track changes while certain important data points may not be sampled due to various reasons, especially on urban road networks. Note that slightly changing the lines has proven useful in other disciplines \cite{Mitchell:MiniRevisited}.


\stitle{Patch points}. Let $\mathcal{R}_{i-1}$, $\mathcal{R}_{i}$ and $\mathcal{R}_{i+1}$ be three continuous directed line segments, where $\mathcal{R}_{i}$ is anomalous, \ie $\mathcal{R}_{i}$ represents only two points. The patch point $G$ \wrt $\mathcal{R}_{i}$ is the intersection point between line segments $\mathcal{R}_{i-1}$ and $\mathcal{R}_{i+1}$.

We next illustrate the use of patch points for reducing anomalous line segments with an example.

\begin{example}
\label{exm-patchpoint}
Consider Figure~\ref{fig:onroad}(2), where $G_1, G_2$ are the patch points \wrt $\mathcal{R}_4$ and $\mathcal{R}_8$, respectively.
After $G_1$ and $G_2$ are interpolated, both algorithms \dpa and \operb return three directed line segments
$\{\vv{P_0G_1}$, $\vv{G_1G_2}$, $\vv{G_2P_{10}}\}$, instead of four and five, respectively, as shown in Example~\ref{exm-anomalous-ls}.
\end{example}

\vspace{-1ex}
\stitle{{Patching method}}. Consider any three continuous directed line segments, $\mathcal{R}_{i-1}=\vv{P_sP_{s+i-1}}$, $\mathcal{R}_{i} = \vv{P_{s+i-1}P_{s+i}}$ and $\mathcal{R}_{i+1}$ $=$ $\vv{P_{s+i}P_{t}}$ such that $\mathcal{R}_{i}$ is anomalous and point $P_{t}$ is an active point and $t>s+i$.
With the practical restrictions and one-pass requirement, the patch point $G$ \wrt $\mathcal{R}_{i}$ needs to satisfy the conditions below, as illustrated by Figure~\ref{fig:patchpoint}.

\sstab(1) Lying on the lines of  $\vv{P_sP_{s+i-1}}$ (\ie $\vv{P_sG}.\theta = \vv{P_sP_{s+i-1}}.\theta$) and  $\vv{P_{s+i}P_{t}}$ (\ie $\vv{GP_{s+i}}.\theta = \vv{P_{s+i}P_{t}}.\theta$),

\sstab(2) $|\vv{P_sG}| \ge (|\vv{P_sP_{s+i-1}}| - \zeta/2)$, and

\sstab(3) the included angle from $\mathcal{R}_{i-1}$ to $\mathcal{R}_{i+1}$ falls in $(-2\pi, -\pi-\gamma_m]$, $[\gamma_m - \pi, \pi-\gamma_m]$ and $[\pi+\gamma_m, 2\pi)$, where $\gamma_m \in [0, \pi]$ is a parameter with $\gamma_m = \frac{\pi}{3}$ by default.

\begin{figure}[tb!]
\centering
\includegraphics[scale = 0.7]{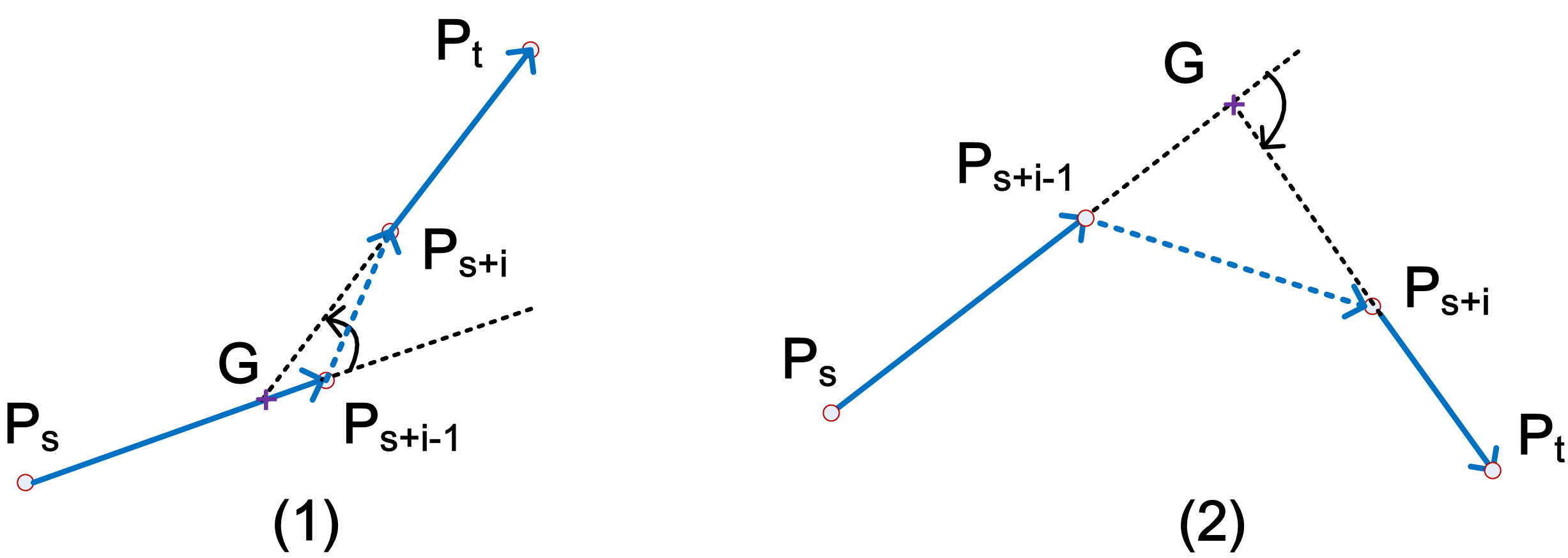}
\vspace{-2ex}
\caption{\small Practical restrictions on patch points.}
\label{fig:patchpoint}
\vspace{-3.5ex}
\end{figure}

\eat{
\stitle{Patch method}. Considering practical restrictions and the one-pass processing requirement, given any two line segments $\mathcal{R}_k = \vv{P_sP_{s+k}}$ and $\mathcal{R}_t = \vv{P_{s+k+1}P_t}$ starting from $P_{s+k+1}$, the patch point $G$ \wrt $\vv{P_{s+k}P_{s+k+1}}$ need to satisfy the following conditions, as illustrated by Figure~\ref{fig:patchpoint}.

\sstab(1) Lying on the lines of $\vv{P_sP_{s+k}}$ (\ie $\vv{P_sG}.\theta = \vv{P_sP_{s+k}}.\theta$) and  $\vv{P_{s+k+1}P_{t}}$ (\ie $\vv{GP_{s+k+1}}.\theta = \vv{P_{s+k+1}P_{t}}.\theta$),

\sstab(2) $|\vv{P_sG}| \ge (|\vv{P_sP_{s+k}}| - \zeta/2)$, and

\sstab(3) the included angle from $\mathcal{R}_k=\vv{P_sP_{s+k}}$  to $\vv{GP_{s+k+1}}$ falls in $(-2\pi, -\pi-\gamma_m]$, $[\gamma_m - \pi, \pi-\gamma_m]$ and $[\pi+\gamma_m, 2\pi)$, where $\gamma_m \in [0,\pi]$ is a parameter with $\gamma_m = \frac{\pi}{3}$ by default.

\begin{figure}[tb!]
\centering
\includegraphics[scale = 0.75]{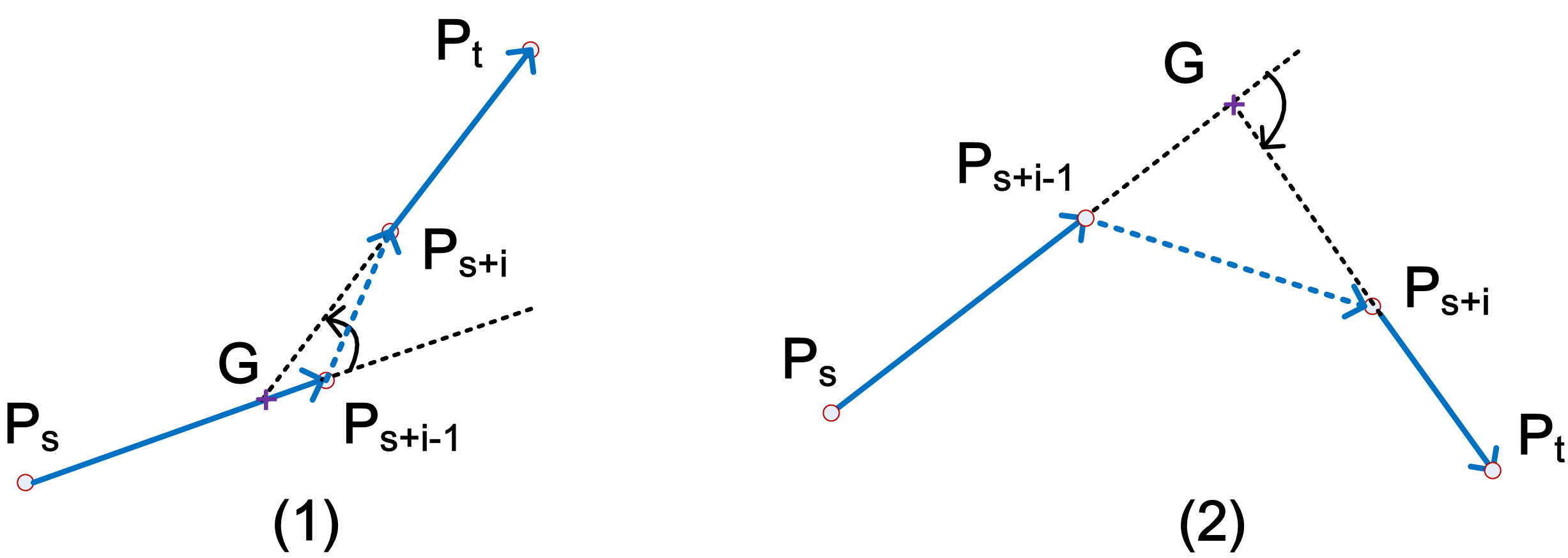}
\caption{Practical restrictions on patch points.}
\label{fig:patchpoint}
\end{figure}
}

Intuitively, these conditions are to incorporate the sudden changes of moving directions implied in a trajectory.
Note that the restriction of the included angles may reduce the chance of eliminating anomalous line segments. However, it helps to produce more rational results.

\subsection{Algorithm OPERB--A}
\label{subsec-operba}
We now present our algorithm \operba that extends algorithm \operb by introducing patch points.
Recall that algorithm \operb starts a new directed line segment when the distance of a point, say $P_{i}$, to the line segment $\mathcal{L}_{i-1}$ is larger than $\zeta/2$, marks the last active point $P_{a}$ as $P_{s+i-1}$, outputs the directed line segment $\mathcal{R}_{i-1} = \vv{P_sP_{s+i-1}}$, and marks $P_{s+i-1}$  as the new start point $P_{s}$ of the remaining subjectify $\dddot{\mathcal{T}_s}[P_{s+i-1}, \ldots, P_{n}]$.
However, for \operba, the line segment $\mathcal{R}_{i-1}$ cannot be outputted until the patch point $G$ is determined when $R_i= \vv{P_{s+i-1}P_{s+i}}$ is an \anoline, and patch point $G$ cannot be determined unless the angle of $\mathcal{R}_{i+1} = \vv{P_{s+i}P_{t}}$ has been determined. Hence, different from \operb, \operba uses a lazy output policy.

\stitle{{The lazy output policy}}. \operba temporarily saves a line segment in memory before outputting it, as follows:

\sstab(1) For simplicity, suppose that the line segment $\mathcal{R}_{i-1}$ is not  anomalous and cannot be compressed with any more points. \operba saves it in memory first.

\sstab(2) It then compresses the subsequent points as \operb to the next line segment $\mathcal{R}_{i}$ until a broken condition is triggered.
If $\mathcal{R}_{i}$ is not anomalous, then it outputs $\mathcal{R}_{i-1}$ and saves $\mathcal{R}_{i}$ in memory.
Otherwise, it marks $\mathcal{R}_{i}$ anomalous, saves it, and moves to the next line segment $\mathcal{R}_{i+1}$.

\sstab(3) If $\mathcal{R}_{i+1}$ is determined and $\mathcal{R}_{i}$ is anomalous, \operba checks the possibility of patching a point $G$ \wrt $\mathcal{R}_{i}$. If so, it outputs $\vv{P_sG}$, and $\vv{GP_t}$ is temporarily saved; Otherwise, it outputs $\mathcal{R}_{i-1}$ and $\mathcal{R}_{i}$, and $\mathcal{R}_{i+1}$ is temporarily saved.

\sstab(4) The process repeats until all points have been processed.


\stitle{Remarks}. All the optimization techniques in Section \ref{subsec-optimizations} remain intact, and are seamlessly integrated into \operba.

We next explain algorithm \operba with an example.

\begin{example}
\label{exam-running-operba}
Algorithm  \operba takes as input the trajectory and $\zeta$ shown in Figure~\ref{fig:dp}, and its output is illustrated in Figure~\ref{exa-fig:operba}.
The process of  \operba is as follows.

\sstab(1) It first creates $\mathcal{L}_0= \mathcal{R}_0 = \vv{P_0P_0}$, compresses $P_1$, \ldots, $P_5$ in turn, and generates $\mathcal{L}_5$, along the same lines as \operb.

\sstab(2) It then finds that $d(P_6, \mathcal{L}_5) > \zeta/2$, which means that $P_6$ cannot be compressed into $\mathcal{R}_5 = \vv{P_0P_5}$. $\mathcal{R}_5$ is temporally saved. And the next line segment starts from $P_5$.

\sstab(3) It then finds that $d(P_7, \mathcal{L}_6) > \zeta/2$ and $\mathcal{R}_6 = \vv{P_5P_6}$ is an \anoline. Hence, $\mathcal{R}_6$ is also temporally saved, and the next line segment starts from $P_6$.

\sstab(4) It continues to compress points $P_7$ and $P_8$ in turn, and generates $\mathcal{L}_8$ along the same lines as \operb.

\sstab(5) It then finds that $d(P_9, \mathcal{L}_8)) > \zeta/2$, which means that $\mathcal{R}_8 = \vv{P_6P_8}$ is determined.
Now \operba tries to expand the lines of $\mathcal{R}_5$ and $\mathcal{R}_8$ to get the intersection point $G$ of them, and uses it as the final start point of $\mathcal{R}_8$.
At last, $\mathcal{R}_5$ is extended to $\vv{P_0G}$ and output as a directed line segment in the result,
$\vv{GP_8}$ is temporarily saved and $P_8$ becomes the start point of the next line segment.

\sstab(6) The above process repeats until all points have been processed. Finally, \operba outputs four line segments $\{\vv{P_0G}$, $\vv{GP_{8}}$, $\vv{P_{8}P_{10}}$, $\vv{P_{10}P_{14}}\}$.

\begin{figure}[tb!]
\centering
\includegraphics[scale = 0.64]{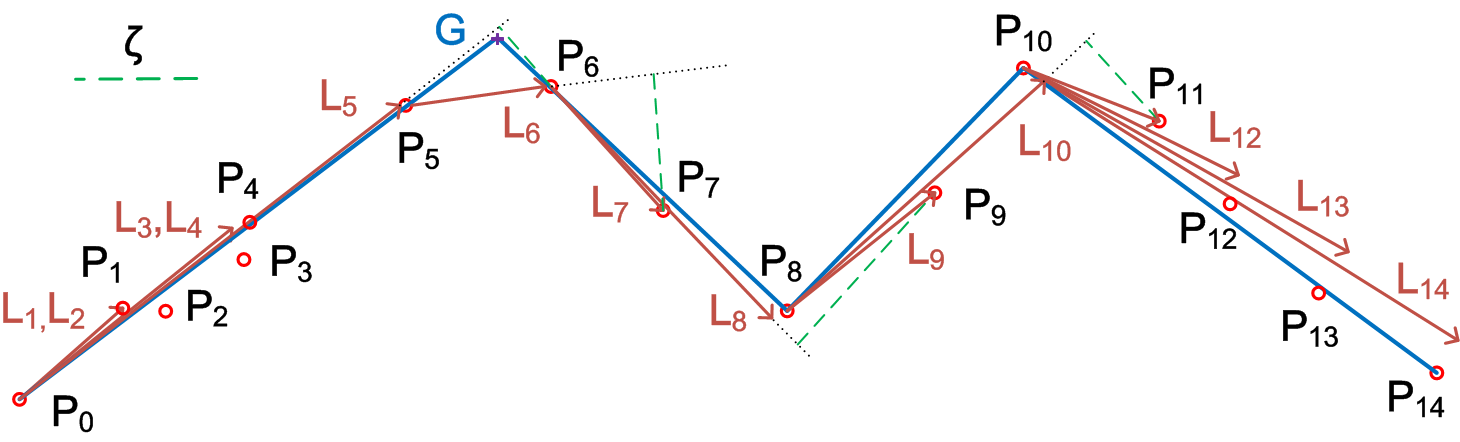}
\vspace{-2.5ex}
\caption{{\small A running example of \operba}. }
\label{exa-fig:operba}
\vspace{-3ex}
\end{figure}

As shown in Figure~\ref{exa-fig:operba}, algorithm \operba further eliminates the directed line segment $\vv{P_5P_6}$, compared with the result of \operb shown in Figure~\ref{exa-fig:operb}.
\end{example}

\vspace{-1ex}
\stitle{Correctness \& complexity analysis}.
Observe that algorithm \operba does not change the angle of any directed line segment compared with \operb, and hence it remains error bounded.
Moreover, each data point in a trajectory is read once and only once in \operba.
Therefore, algorithm \operba is one-pass and error bounded.
It is also easy to verify that algorithm \operba takes $O(1)$ space, the same as \operb, as the directed line segment in $\overline{\mathcal{T}}$ can be outputted immediately once it is generated.
That is, the nice properties of \operb remain intact in \operba.


\section{Experimental Study}
\label{sec-exp}
In this section, we present an extensive experimental study of our one-pass error bounded algorithms \operb and \operba.
Using four real-life datasets, we conducted four sets of experiments to evaluate:
(1) the execution time of our approaches compared with algorithms \dpa and \fbqsa, and the impacts of optimizations,
(2) the compression ratios of our approaches compared with \dpa and \fbqsa, and the impacts of optimizations,
(3) the average errors of our approaches compared with algorithms \dpa and \fbqsa, and
(4) the effectiveness of trajectory interpolation.

\subsection{Experimental Setting}

\stitle{Real-life Trajectory Datasets}.
We use four real-life datasets shown in Table~\ref{tab:dataset} to test our solutions.

\sstab{\bf(1) Taxi trajectory data}, referred to as \taxi, is the GPS trajectories collected by $12,727$ taxies equipped with GPS sensors in Beijing during a period
from Nov. 1, 2010 to Nov. 30, 2010. The sampling rate was one point  per 60s, and \taxi has $39,100$ data points on average per trajectory.

\sstab{\bf(2) Truck trajectory data}, referred to as \truck, is the GPS trajectories collected by 10,368 trucks equipped with GPS sensors in China
during a period from Mar. 2015 to Oct. 2015. The sampling rate varied from 1s to 60s. Trajectories mostly have around $50$ to $90$ thousand data points.

\sstab{\bf(3) Service car trajectory data}, referred to as \sercar,  is the GPS trajectories collected by a car rental company.
We chose $11,000$ cars from them, during Apr. 2015 to Nov. 2015. The sampling rate was one point per $3$--$5$ seconds, and
each trajectory has around $119.1K$ data points.

{\sstab{\bf(4) GeoLife trajectory data}, refered to as \geolife, is the GPS trajectories collected in GeoLife project~\cite{Zheng:GeoLife} by 182 users in a period from Apr. 2007 to Oct. 2011. These trajectories have a variety of sampling rates, among which 91\% are logged in each 1-5 seconds or each 5-10 meters per point. The longest trajectory has 2,156,994 points.}


\begin{table}
\vspace{1.5ex}
\centering
\small
\begin{tabular}{|l|c|c|c|r|}
\hline
\kw{Data}& \kw{Number\ of}     &\kw{Sampling}   &\kw{Points Per}    &\kw{Total} \\
\kw{Sets} & \kw{Trajectories}   &\kw{Rates (s)}  &\kw{Trajectory (K)}&\kw{points}\\
\hline\hline
\taxi	&12,727	    &60	        &$\sim39.1$      &498M \\
\hline
\truck	&10,368	    &1-60	    &$\sim71.9$     &746M \\
\hline
\sercar	&11,000	    &3-5	    &$\sim119.1$   &1.31G\\
\hline
\geolife &182	    &1-5	    &$\sim132.8$   &24.2M\\
\hline
\end{tabular}
\vspace{-1ex}
\caption{\small Real-life trajectory datasets}\label{tab:dataset}
\vspace{-4ex}
\end{table}


\stitle{Algorithms and implementation}.
We compared our algorithms \operb and \operba with two existing \lsa algorithms \dpa \cite{Douglas:Peucker}  and \fbqsa \cite{Liu:BQS}, and algorithms \kw{Raw}-\operb and \kw{Raw}-\operba,
the counterparts of \operb and \operba without optimizations, respectively.

\sstab{(1) Algorithm \dpa} is a classic batch \lsa algorithm with an excellent compression ratio (shown in Figure~\ref{alg:dp}).

\sstab{(2) Algorithm  \fbqsa} is an online algorithm, and is the fastest existing \lsa algorithm (recall Section~\ref{subsec-lsalg}).

\sstab{(3) Algorithm  \operb} combines the algorithm in Figure~\ref{alg:operb} and the optimization techniques in Section~\ref{subsec-optimizations}, while
algorithm \kw{Raw}-\operb is the basic algorithm in Figure~\ref{alg:operb} only.

\sstab{(4) Algorithms \operba and \kw{Raw}-\operba} are the aggressive solutions extending \operb and \kw{Raw}-\operb with trajectory interpolation, respectively.

All algorithms were implemented with Java.
All tests were run on an x64-based  PC with 4 Intel(R) Core(TM) i5-4570 CPU @ 3.20GHz  and 16GB of memory, and each test was repeated
over 3 times and the average is reported here.

\begin{figure*}[tb!]
\centering
\includegraphics[scale = 0.47]{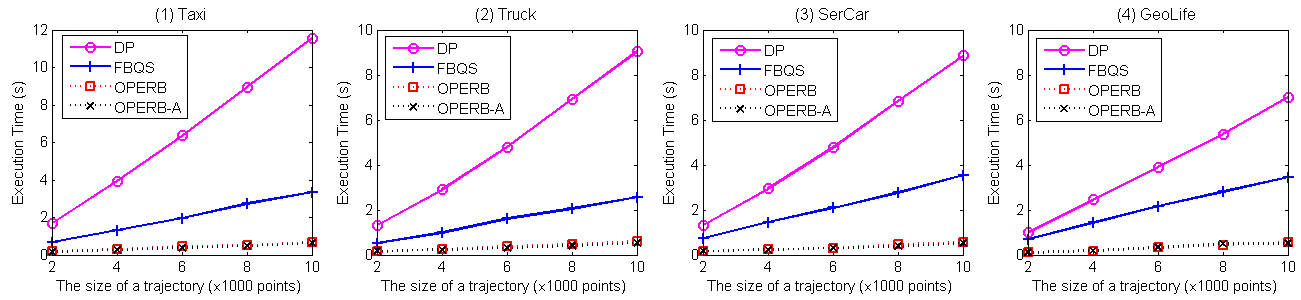}
\vspace{-2.5ex}
\caption{\small Efficiency evaluation: varying the size of trajectories.}\label{fig:time-size}
\vspace{-1ex}
\end{figure*}
\begin{figure*}[tb!]
\centering
\includegraphics[scale = 0.47]{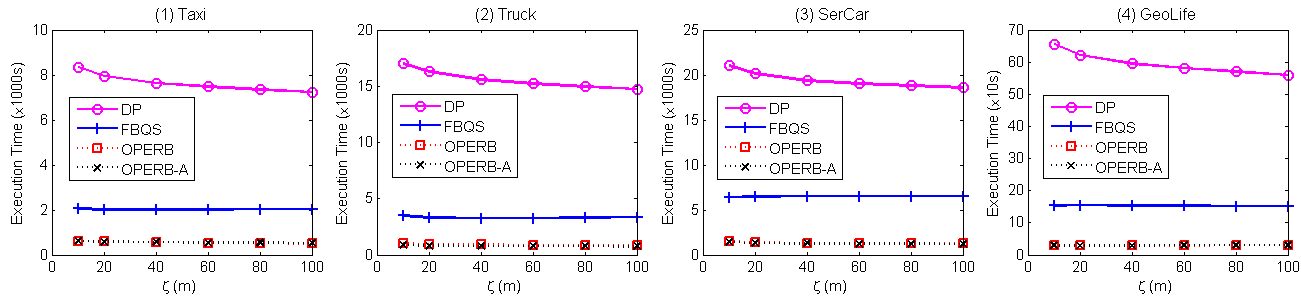}
\vspace{-2.5ex}
\caption{\small Efficiency evaluation: varying the error bound $\zeta$.}
\label{fig:time-zeta}
\vspace{-1ex}
\end{figure*}
\begin{figure*}[tb!]
\centering
\includegraphics[scale = 0.47]{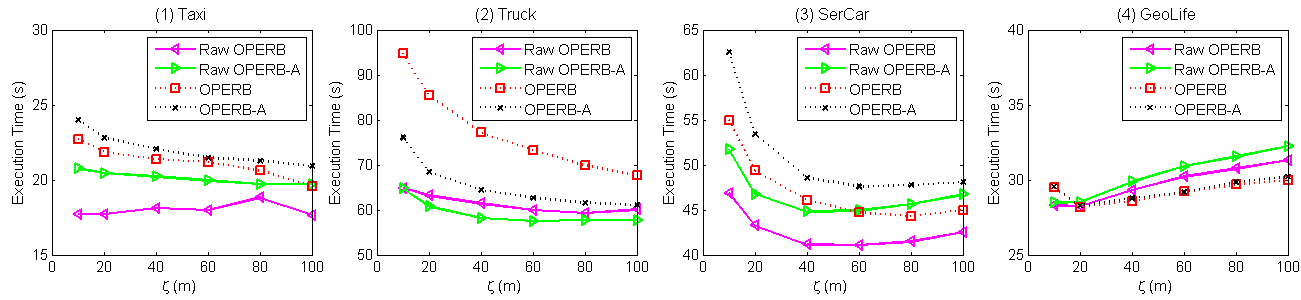}
\vspace{-2.5ex}
\caption{\small Efficiency evaluation: optimization techniques when varying the error bound $\zeta$.}
\label{fig:time-opt}
\vspace{-2ex}
\end{figure*}

\subsection{Experimental Results}

\subsubsection{Evaluation of Compression Efficiency}
In the first set of tests, we compare the efficiency (execution time) of our approaches \operb and \operba with algorithms \dpa and \fbqsa
and with algorithms \kw{Raw}-\operb and \kw{Raw}-\operba.
For fairness, we load and compress trajectories one by one, and only count the running time of the compressing process.

\stitle{{Exp-1.1}: Impacts of the sizes of trajectories}.
To evaluate the impacts of the number of data points in a trajectory (\ie the size of a trajectory),
we chose $100$ trajectories from \taxi, \truck, \sercar and \geolife, respectively,
and varied the size \trajec{|T|} of trajectories from $2,000$ to $10,000$, while fixed $\zeta = 40$ meters (m).
The results are reported in Figure~\ref{fig:time-size}.

\sstab(1) Algorithms \operb, \operba and \fbqsa  scale well with the increase of the size of trajectories on all datasets,
and show a linear running time, while algorithm \dpa does not.
This is consistent with their time complexity analyses.

\sstab(2) Algorithms \operb and \operba are the fastest \lsa algorithms, and are {($3.8$--$5.3$, $3.5$--$4.8$, $4.6$--$7.2$, $6.2$--$8.4$)} times faster than \fbqsa,
and {($9.6$--$17.6$, $8.8$--$15.4$, $8.4$--$16.3$, $9.0$--$14.4$)} times faster than \dpa on (\taxi, \truck, \sercar, \geolife), respectively. The running time of \operb and \operba is similar, and the difference is below 10\%.
\eat{Moreover, \operba is a bit faster than \operb,
especially when there are more than $1,000$ data points in  trajectories.}

\stitle{{Exp-1.2}:  Impacts of the error bound $\zeta$}.
To evaluate the impacts of $\zeta$, we varied $\zeta$ from $10m$ to $100m$ on the entire \taxi, \truck, \sercar and \geolife, respectively.
The results are reported in Figure~\ref{fig:time-zeta}.


\sstab(1) All algorithms are not very sensitive to $\zeta$, but their running time all decreases a little bit with the increase of $\zeta$,
as the increment of $\zeta$ decreases the number of directed line segments in the output.
Further, algorithm \dpa is more sensitive to $\zeta$ than the other three algorithms.

\sstab(2) Algorithms \operb and \operba are obviously faster than \dpa and \fbqsa in all cases.
\operb is on average ($13.9$, $17.4$, $14.7$, {$20.6$}) times faster than \dpa, and ($4.1$, $4.1$, $5.4$, {$5.2$}) times faster than \fbqsa on (\taxi, \truck, \sercar, {\geolife}), respectively. Algorithm \operba is as fast as \operb because trajectory interpolation is a light weight operation.

\stitle{Exp-1.3: Impacts of optimization techniques}.
To evaluate the impacts of our optimization techniques (see Section~\ref{subsec-optimizations}),
we compared algorithms \operb and \operba with \kw{Raw}-\operb and \kw{Raw}-\operba, respectively.
We chose $500$, $1000$ and $500$ trajectories from \taxi, \truck, \sercar and \geolife, respectively, and varied $\zeta$ from 10m to 100m.
The results are reported in Figure~\ref{fig:time-opt}.

\sstab(1) The running time of all algorithms slightly decreases with the increase of $\zeta$, consistent with {Exp-1.2}.

\begin{figure*}[tb!]
\centering
\includegraphics[scale=0.47]{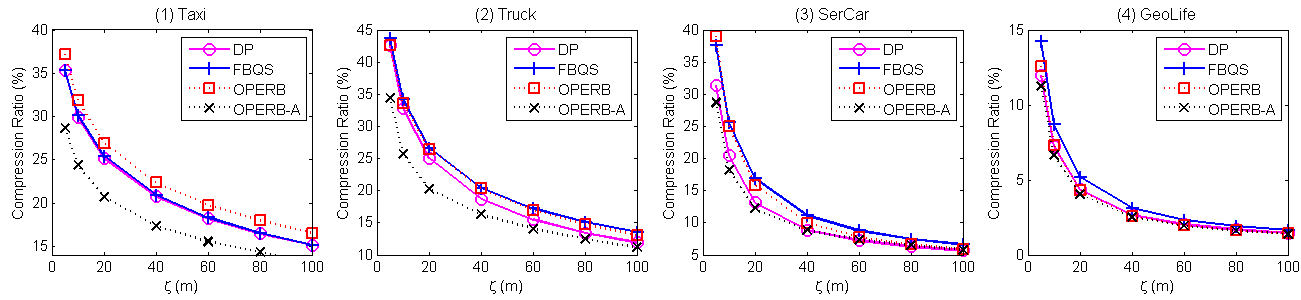}
\vspace{-2.5ex}
\caption{\small Effectiveness evaluation: varying the error bound $\zeta$.}
\label{fig:cr}
\vspace{-1ex}
\end{figure*}
\begin{figure*}[tb!]
\centering
\includegraphics[scale=0.47]{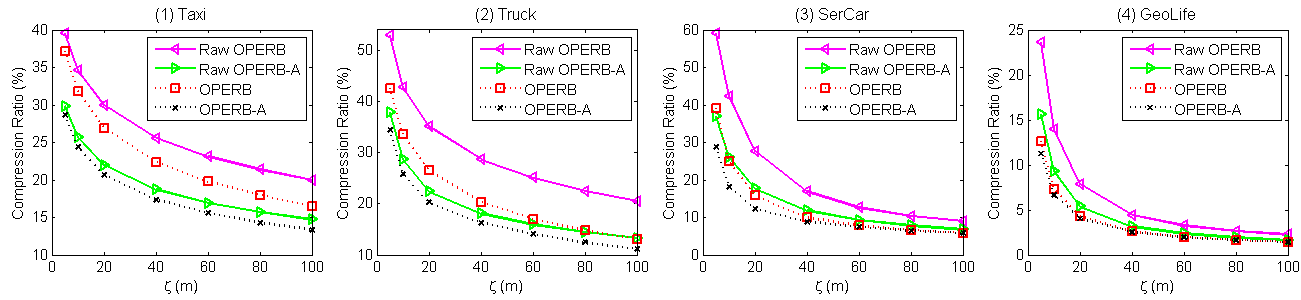}
\vspace{-2.5ex}
\caption{\small Effectiveness evaluation: optimization techniques when varying the error bound $\zeta$.}
\label{fig:optim}
\vspace{-2.5ex}
\end{figure*}

\sstab(2) The running time of \kw{Raw}-\operb is {$(85.0\%$, $79.6\%$, $90.4\%$, $100.4\%$)} of \operb on average on ($\taxi$, $\truck$, $\sercar$, $\geolife$), respectively,
and the running time of \kw{Raw}-\operba is {$(91.3\%$, $90.1\%$, $91.6\%$, $101.5\%$) }of \operba on average on ($\taxi$, $\truck$, $\sercar$, $\geolife$), respectively.
This shows that the optimization techniques have a limited impact on the efficiency of \operb and \operba.
However, as will be shown immediately, the benefits of compression ratios are highly appreciated.

\subsubsection{Evaluation of Compression Effectiveness}

In the second set of tests, we compare the compression ratios of our algorithms \operb and \operba with \dpa and \fbqsa
and with \kw{Raw}-\operb and \kw{Raw}-\operba, respectively.
Given a set of trajectories $\{\dddot{\mathcal{T}_1}, \ldots, \dddot{\mathcal{T}_M}\}$ and their piecewise line representations
$\{\overline{\mathcal{T}_1}, \ldots, \overline{\mathcal{T}_M}\}$,
 the compression ratio is $(\sum_{j=1}^{M} |\overline{\mathcal{T}}_j |)/(\sum_{j=1}^{M} |\dddot{\mathcal{T}}_j |)$.
 Note that by the definition, algorithms with lower compression ratios are better.

\stitle{Exp-2.1: Impacts of the error bound $\zeta$}.
To evaluate the impacts of $\zeta$ on compression ratios of these algorithms, we varied $\zeta$ from $5m$ to $100m$ on
 the entire four datasets, respectively.
The results are reported in Figure~\ref{fig:cr}.

\sstab (1) When increasing $\zeta$, the compression ratios decrease. For example, in \sercar,
the compression ratios are greater than $28.7\%$ when $\zeta$ = $5m$, but are less than $6.6\%$ when $\zeta$ = $100m$.

\sstab (2) \geolife has the lowest compression ratios, compared with \taxi, \truck and \sercar,
due to its highest sampling rate, \taxi has the highest compression ratios due to its lowest sampling rate, and \truck and \sercar have the compression ratios in the middle accordingly.

\sstab (3) {First, algorithm \operb has comparable compression ratios with \fbqsa and \dpa.
For example, when $\zeta = 40m$, the compression ratios are ($20.9\%$, $20.4\%$, $11.1\%$, $3.1\%$) of \fbqsa, ($20.7\%$, $18.7\%$, $8.9\%$, $2.6\%$) of \dpa and ($22.3\%$, $20.3\%$, $10.0\%$, $2.6\%$) of \operb on (\taxi, \truck, \sercar, \geolife), respectively.
For all $\zeta$, the compression ratios of \operb are on average ($107.2\%$, $98.3\%$, $92.9\%$, {$85.1\%$}) of \fbqsa and ($107.7\%$, $106.6\%$, $113.5\%$, $99.6\%$) of \dpa on (\taxi, \truck, \sercar, \geolife), respectively.
\operb is better than  \fbqsa on \truck, \sercar and \geolife, while a little worse on \taxi. The results also show that \operb has a better performance than \fbqsa on datasets with high sampling rates.}

\ni Second, algorithm \operba achieves the best compression ratios on all datasets and nearly all $\zeta$ values.
Its compression ratios are on average ($83.7\%$, $79.5\%$, $79.7\%$, $81.0\%$) of \fbqsa and ($84.2\%$, $86.4\%$, $97.1\%$, $94.7\%$) of \dpa on (\taxi, \truck, \sercar, \geolife), respectively.
Similar to \operb, \operba has advantages on datasets with high sampling rates.

\stitle{{Exp-2.2}: Impacts of the optimization techniques}.
We compared algorithms \operb and \operba with \kw{Raw}-\operb and \kw{Raw}-\operba, respectively.
We varied $\zeta$ from $5m$ to $100m$ on the entire \taxi, \truck, \sercar and \geolife, respectively.
The results are reported in Figure~\ref{fig:optim}.

\begin{figure*}[tb!]
\centering
\includegraphics[scale=0.47]{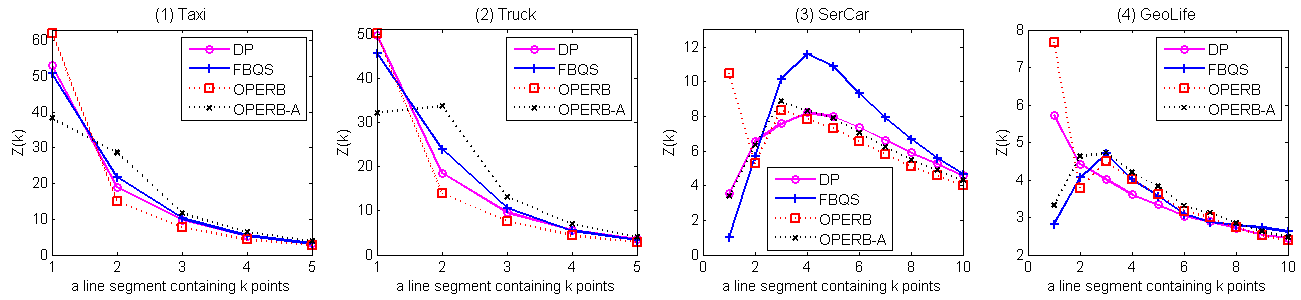}
\vspace{-2.5ex}
\caption{\small Effectiveness evaluation: distribution of line segments. The horizontal coordinate is the number $k$ of data points,
and the vertical coordinate is the number $\mathcal{Z}(k)$ of line segments containing $k$ points.}
\label{fig:dop}
\vspace{-1ex}
\end{figure*}

\begin{figure*}[tb]
\centering
\includegraphics[scale = 0.47]{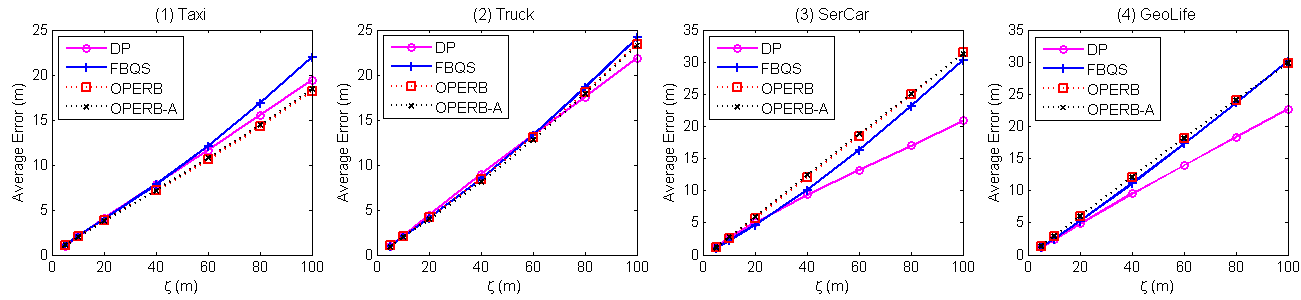}
\vspace{-2.5ex}
\caption{\small Evaluation of average errors: varying the error bound $\zeta$.}
\label{fig:ae}
\vspace{-3ex}
\end{figure*}

\sstab(1) The optimizations have great impacts on the compression ratios of \operb and \operba. Indeed,
\operb is  on average {($87.9\%$, ${71.8\%}$, $61.8\%$, $58.0\%$)} of \kw{Raw}-\operb, and \operba is  on average {($93.1\%$, ${88.5\%}$, $77.1\%$, $78.5\%$)} of \kw{Raw}-\operba
on (\taxi, \truck, \sercar, \geolife), respectively. Note that the optimization techniques  have a better impact on datasets with high sampling rates.

\sstab(2) The impacts of the optimization techniques increase with the increase of $\zeta$ on \taxi and \truck.
For example, \operb is {($92.1\%$, ${87.3\%}$, $82.6\%$)} of \kw{Raw}-\operb on \taxi, and {($78.5\%$, $70.9\%$, $63.7\%$)} of \kw{Raw}-\operb on \truck
when $\zeta$ = ($10$, $40$, $100$), respectively. It is similar for algorithm \operba.

\stitle{{Exp-2.3}: Distribution of line segments}.
To further evaluate the difference of the compression results of these algorithms, we chose $100$ trajectories from each of (\taxi, \truck, \sercar, \geolife),
while fixed $\zeta=40m$.
For a $\overline{\mathcal{T}}$ = $(\mathcal{L}_1, \mathcal{L}_2, \cdots , \mathcal{L}_M)$ derived
from trajectories by a compression algorithm, we count the number of points, $C_i$, included in each $\mathcal{L}_i$,
then let $\mathcal{Z}(k) = |\{ C_i | C_i =k\}|$, \ie $\mathcal{Z}(5)$ means the number of all $\mathcal{L}_i$ containing 5 data points.
Note that the start/end points are repeatedly counted for adjacent line segments, and, hence, there may be some lines having only one point.
The results are reported in Figure~\ref{fig:dop}.


\sstab(1) Algorithms \operba and \dpa produce more line segments containing large number of points (heavy line segments) than \fbqsa and \operb.
Heavy line segments are closely related to compression ratios, and help to decrease compression ratios.
The distribution of line segments is consistent with the compression ratios shown above.

\sstab(2) {Algorithm \operb has the largest number of line segments containing only one point. However, they are reduced by \operba to a large extent.}

\vspace{-0.5ex}
\subsubsection{Evaluation of Average Errors}
In the third set of tests, we evaluate the average errors of these algorithms.
We varied $\zeta$ from $5m$ to $100m$ on the entire \taxi, \truck, \sercar and \geolife, respectively.
Given a set of trajectories $\{\dddot{\mathcal{T}_1}, \ldots, \dddot{\mathcal{T}_M}\}$ and their piecewise line representations
$\{\overline{\mathcal{T}_1}, \ldots, \overline{\mathcal{T}_M}\}$, and point $P_{j,i}$ denoting
a point in trajectory $\dddot{\mathcal{T}}_j$ contained in a line segment $\mathcal{L}_{l,i}\in\overline{\mathcal{T}_l}$ ($l\in[1,M]$),
then the average error is $\sum_{j=1}^{M}\sum_{i=0}^{M} d(P_{j,i},
\mathcal{L}_{l,i})/\sum_{j=1}^{M}{|\dddot{\mathcal{T}}_j |}$.
The results are reported in Figure~\ref{fig:ae}.

\sstab(1) Average errors obviously increase with the increase of $\zeta$. \taxi also has the lowest average errors than \truck and \sercar for all algorithms,
as it has the highest compression ratios, and \sercar has the highest average errors, as it has the lowest compression ratios among the three datasets.

\sstab(2) Algorithm \dpa not only has better compression ratios, but also has lower average errors than algorithm \fbqsa on all datasets and most $\zeta$ values.

\sstab(3) Algorithms \operb and \operba have similar average errors with each other. Meanwhile, \operba does not introduce extra error. Compared with \dpa and \fbqsa, their errors are a litter smaller on \taxi and a little larger on \sercar.

\vspace{-0.5ex}
\subsubsection{Evaluation of Trajectories Interpolation}

In the last set of tests, we evaluate the lazy trajectory interpolation policy of algorithm \operba.

\begin{figure}[tb!]
\centering
\includegraphics[scale=0.465]{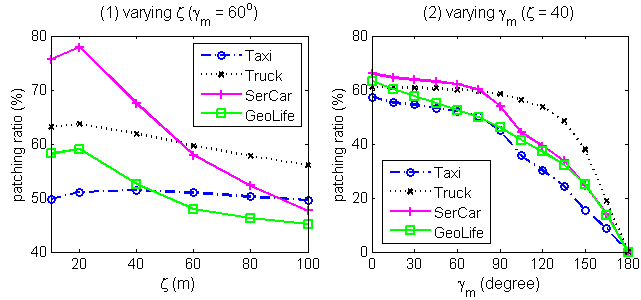}
\vspace{-2ex}
\caption{\small Patching ratios of \operba.}
\label{fig:patch}
\vspace{-4ex}
\end{figure}

\stitle{{Exp-4.1}: Patching ratios}.
To evaluate the patching ratios of \operba and the impacts of $\zeta$ on patching ratios, we varied $\zeta$ from $10$m to $100$m, while fixed $\gamma_m = \pi/3$,
on \taxi, \truck, \sercar and \geolife, respectively.
The patching ratio is defined as $\frac{N_p} {N_a} * 100\%$, where $N_p$ is the number of patching points successful added to the trajectories,
and $N_a$ is the number of anonymous line segments before trajectory interpolation. The results are reported in Figure~\ref{fig:patch}--(1).

\sstab(1) The patching ratios are on average {($50.5\%$, $60.3\%$, $63.2\%$, $51.5\%$)} on (\taxi, \truck, \sercar, \geolife), respectively.
That is, more than half of the anomalous line segments are successfully eliminated by the lazy trajectory output policy.

\sstab(2) {The patching ratios on \taxi, \truck, \sercar and \geolife decrease from $\zeta=30m$ or $\zeta=40m$, respectively}.

\sstab(3) \taxi has the lowest patching ratios due to its relatively lower sampling rate. Moreover, the number of anomalous line segments output by \operba
is significantly less than the other algorithms.
That is, even a patching ratio like  {50.5\%} is enough to improve the compression ratio and to make \operba have the best compression ratio.

\stitle{Exp-4.2: Impacts of $\gamma_m$ for trajectory interpolation}.
Our trajectory interpolation uses a parameter $\gamma_m \in [0,\pi]$ to restrict the included angle of two line segments.
Note that a smaller $\gamma_m$ means that a larger direction change is allowed in trajectory interpolation.
 To evaluate the impacts of $\gamma_m$ on patching ratios, we randomly chose $100$ trajectories from each of (\taxi, \truck, \sercar),
and we varied $\gamma_m$ from $0^o$ to $180^o$ (or 0 to $\pi$),  while fixed $\zeta =40m$.
The results are reported Figure~\ref{fig:patch}--(2).

\sstab(1) When varying $\gamma_m$, the patching ratio decreases with the increase of $\gamma_m$.
The patching ratio decreases (a) slowly when $\gamma_m \in [0,75^o]$, (b) fast when $\gamma_m \in (75^o, 145^o)$,
and (c) fastest when $\gamma_m \in (145^o, 180^o]$. The region $[30^o, 90^o]$ is the candidate region for $\gamma_m$
with both high patching ratios and reasonable patch points, not too far away from the points $P_{s+k}$
and $P_{s+k+1}$, between which the patch point is interpolated. Hence, we set $\gamma_m = \pi/3$ by default.

\sstab(2) Parameter $\gamma_m$ has different impacts on different data sets. The patching ratio of \taxi decreases quickly
when $\gamma_m \in [80^o, 120^o]$, it is the result of taxies running on urban road networks, which have more crossroads.
The patching ratio of \truck decreases slowly in this region because many trucks are running in suburban districts or between cities, in which there are
less crossroads.

\stitle{Summary}.
From these tests we find the following.

\emph{\sstab{(1) Efficiency}}. \operb and \operba are the fastest algorithms, which are on average $(13.9, ~17.4, ~14.7, {~20.6})$ times faster than \dpa, and $(4.1,~4.1,~5.4, {~5.2})$
times faster than \fbqsa on (\taxi, \truck, \sercar, \geolife), respectively.

\emph{\sstab{(2) Compression ratios}}. (a) \operb is comparable {with \fbqsa and \dpa}. Its compression ratios are on average $(107.2\%, ~98.3\%, ~92.9\%, ~85.1\%)$ of \fbqsa and ($107.7\%$, $106.6\%$, $113.5\%$, $99.6\%$) of \dpa on (\taxi, \truck, \sercar, \geolife), respectively, and \operb has a better performance on trajectories with higher sampling rates.
(b) \operba has the best compression ratios on all datasets and nearly all $\zeta$ values.
Its compression ratios are on average {($83.7\%$, $79.5\%$, $79.7\%$, $81.0\%$)} of \fbqsa and {($84.2\%$, $86.4\%$, $97.1\%$, $94.7\%$)} of \dpa on (\taxi, \truck, \sercar, \geolife), respectively.
It shows its advantage on trajectories with both high and low sampling rates.
(c) The optimization techniques are effective for algorithms \operb and \operba on all datasets.

\emph{\sstab{(3) Average errors}}. {Algorithm \operb has similar average errors with \operba. They have lower average errors than the other algorithms on \taxi while higher on \sercar.}

\emph{\sstab{(4) Patching ratios}}. \operba  successfully eliminates more than a half of anomalous line segments by patching data points,
which improves the compression ratio, and indeed makes it achieve the best compression ratio.

\vspace{-0.5ex}
\section{Conclusions and Future Work}
We have proposed \operb and \operba, two one-pass error bounded trajectory simplification algorithms.
First, we have developed a novel local distance checking approach, based on which we then have designed \operb, together with optimization techniques for improving its compression ratio.
Second, by allowing interpolating new data points into a trajectory under certain conditions, we have developed an aggressive one-pass error bounded trajectory simplification algorithm \operba, which has significantly improved the compression ratio.
Finally, we have experimentally verified that both \operaa and \operab are much faster than \fbqsa, the fastest existing \lsa algorithm,
and in terms of compression ratio, \operaa is comparable with \dpa, and \operab is better than \dpa on average, the existing \lsa algorithm with the best compression ratio.

A couple of issues need further study. We are to incorporate semantic based methods and to study alternative forms of fitting functions to further improve the effectiveness of trajectory compression.

%

\stitle{Acknowledgments}.
This work is supported in part by NSFC {\small U1636210}, 973 Program {\small 2014CB340300} \& NSFC {\small 61421003}. 
 For any correspondence, please refer to Shuai Ma.

\bibliographystyle{abbrv}
\begin{small}
\bibliography{opera-ref}
\end{small}


%
%


\end{document}